\documentclass[a4paper,11pt]{article}
\usepackage{amssymb}
\usepackage{amsfonts}
\usepackage{graphicx}
\usepackage{epstopdf}
\usepackage{dcolumn}
\usepackage{amsmath}
\usepackage{latexsym,bm}
\usepackage{geometry}

\def \be {\begin{equation}}
\def \ee {\end{equation}}
\def \bea {\begin{eqnarray}}
\def \eea {\end{eqnarray}}
\def \nn {\nonumber}

\def \a {\alpha}
\def \b {\beta}

\def \d {\delta}

\def \m {\mu}
\def \n {\nu}
\def \k {\kappa}

\def \s {\sigma}
\def \r {\rho}
\def \o {\omega}

\def \th {\theta}
\def \Th {\Theta}

\def \t {\tau}
\def \dag {\dagger}
\def \p {\partial}

\def\bd{\begin{document}}
\def\ed{\end{document}}
\def\nn{\nonumber}
\def\bea{\begin{eqnarray}}
\def\eea{\end{eqnarray}}
\let\bm=\bibitem
\let\la=\label

\def\N{{\cal N}}
\def\sst{\scriptscriptstyle}
\def\thetabar{\bar\theta}
\def\Tr{{\rm Tr}}
\def\one{\mbox{1 \kern-.59em {\rm l}}}

\def\a{\alpha}      \def\da{{\dot\alpha}}
\def\b{\beta}       \def\db{{\dot\beta}}
\def\c{\gamma}  \def\C{\Gamma}  \def\cdt{\dot\gamma}
\def\d{\delta}  \def\D{\Delta}  \def\ddt{\dot\delta}
\def\e{\epsilon}        \def\vare{\varepsilon}
\def\f{\phi}    \def\F{\Phi}    \def\vvf{\f}
\def\h{\eta}
\def\k{\kappa}
\def\l{\lambda} \def\L{\Lambda}
\def\m{\mu} \def\n{\nu}
\def\o{\omega}
\def\P{\Pi}
\def\r{\rho}
\def\s{\sigma}  \def\S{\Sigma}
\def\t{\tau}
\def\th{\theta} \def\Th{\Theta} \def\vth{\vartheta}
\def\X{\Xeta}
\def\z{\zeta}
\def\w{\wedge}
\def\u{\underline}
\def\hs{\hspace}

\def\cA{{\cal A}} \def\cB{{\cal B}} \def\cC{{\cal C}}
\def\cD{{\cal D}} \def\cE{{\cal E}} \def\cF{{\cal F}}
\def\cG{{\cal G}} \def\cH{{\cal H}} \def\cI{{\cal I}}
\def\cJ{{\cal J}} \def\cK{{\cal K}} \def\cL{{\cal L}}
\def\cM{{\cal M}} \def\cN{{\cal N}} \def\cO{{\cal O}}
\def\cP{{\cal P}} \def\cQ{{\cal Q}} \def\cR{{\cal R}}
\def\cS{{\cal S}} \def\cT{{\cal T}} \def\cU{{\cal U}}
\def\cV{{\cal V}} \def\cW{{\cal W}} \def\cX{{\cal X}}
\def\cY{{\cal Y}} \def\cZ{{\cal Z}}

\def\ua{\underline{\alpha}} \def\ubb{\underline{\beta}}
\def\ug{\underline{\gamma}}
\def\ub{\underline{\phantom{\alpha}}\!\!\!\beta}
\def\uc{\underline{\phantom{\alpha}}\!\!\!\gamma}
\def\um{\underline{\mu}} \def\un{\underline{\nu}}
\def\ud{\underline\delta}
\def\ue{\underline\epsilon}
\def\una{\underline a}\def\unA{\underline A}
\def\unb{\underline b}\def\unB{\underline B}
\def\unc{\underline c}\def\unC{\underline C}
\def\und{\underline d}\def\unD{\underline D}
\def\une{\underline e}\def\unE{\underline E}
\def\unf{\underline{\phantom{e}}\!\!\!\! f}\def\unF{\underline F}
\def\unm{\underline m}\def\unM{\underline M}
\def\unn{\underline n}\def\unN{\underline N}
\def\unp{\underline{\phantom{a}}\!\!\! p}\def\unP{\underline P}
\def\unq{\underline{\phantom{a}}\!\!\! q}
\def\unQ{\underline{\phantom{A}}\!\!\!\! Q}
\def\unH{\underline{H}}
\def\ul{\underline}
\def\As {{A \hspace{-6.4pt} \slash}\;}
\def\bs {{b \hspace{-6.4pt} \slash}\;}
\def\Ds {{D \hspace{-6.4pt} \slash}\;}
\def\ds {{\del \hspace{-6.4pt} \slash}\;}
\def\ss {{\s \hspace{-6.4pt} \slash}\;}
\def\ks {{ k \hspace{-6.4pt} \slash}\;}
\def\ps {{p \hspace{-6.4pt} \slash}\;}
\def\pas {{{p_1} \hspace{-6.4pt} \slash}\;}
\def\pbs {{{p_2} \hspace{-6.4pt} \slash}\;}

\def\Fh{\hat{F}}
\def\Vh{\hat{V}}
\def\Xh{\hat{X}}
\def\ah{\hat{a}}
\def\xh{\hat{x}}
\def\yh{\hat{y}}
\def\ph{\hat{p}}
\def\xih{\hat{\xi}}

\def\psit{\tilde{\psi}}
\def\Psit{\tilde{\Psi}}
\def\tht{\tilde{\th}}

\def\At{\tilde{A}}
\def\Qt{\tilde{Q}}
\def\Rt{\tilde{R}}
\def\Nt{\tilde{N}}

\def\at{\tilde{a}}
\def\st{\tilde{s}}
\def\ft{\tilde{f}}
\def\pt{\tilde{p}}
\def\qt{\tilde{q}}
\def\vt{\tilde{v}}
\def\nt{\tilde{n}}

\def\delb{\bar{\partial}}
\def\bz{\bar{z}}
\def\bD{\bar{D}}
\def\bB{\bar{B}}

\def\bk{{\bf k}}
\def\bl{{\bf l}}
\def\bp{{\bf p}}
\def\bq{{\bf q}}
\def\br{{\bf r}}
\def\bx{{\bf x}}
\def\by{{\bf y}}
\def\bR{{\bf R}}
\def\bV{{\bf V}}

\def\d{\delta}\def\D{\Delta}\def\ddt{\dot\delta}

\def\p{\partial} \def\del{\partial}
\def\xx{\times}
\def\uno{\mbox{1 \kern-.59em {\rm l}}}

\def\trp{^{\top}}
\def\inv{^{-1}}
\def\dag{{^{\dagger}}}

\def\pr{\prime}
\def\rar{\rightarrow}
\def\lar{\leftarrow}
\def\lrar{\leftrightarrow}

\title{Conical Defects, Black Holes and Higher Spin (Super-)Symmetry}
\author{
Bin Chen$^{1,2,3}$\footnote{bchen01@pku.edu.cn},\,
Jiang Long$^{1,2}$\footnote{lj301@pku.edu.cn}\,
and
Yi-Nan Wang$^{1}$\footnote{ynwang@pku.edu.cn}
}
\date{}

\begin{document}

\maketitle
\begin{center}
{\it
$^{1}$Department of Physics, Peking University, Beijing 100871, P.R. China\\
\vspace{2mm}
$^{2}$State Key Laboratory of Nuclear Physics and Technology, Peking University, Beijing 100871, P.R. China\\
\vspace{2mm}
$^{3}$Center for High Energy Physics, Peking University, Beijing 100871, P.R. China\\
}
\vspace{10mm}
\end{center}

\begin{abstract}

We study the (super-)symmetries of classical solutions in the higher spin (super-)gravity in AdS$_3$. We show that the symmetries of the solutions are encoded in the holonomy around the spatial circle. When the spatial holonomies of the solutions are trivial, they preserve maximal symmetries of the theory, and are actually the smooth conical defects. We find all the smooth conical defects in the $sl(N), so(2N+1),sp(2N), so(2N), g_2$, as well as in $sl(N|N-1)$ and $osp(2N+1|2N)$ Chern-Simons gravity theories.  In the bosonic higher spin cases, there are one-to-one correspondences between the smooth conical defects and the highest weight representations of Lie group. Furthermore we investigate the higher spin black holes in $osp(3|2)$ and $sl(3|2)$ higher spin (super-)gravity and find that they are only partially symmetric. In general, the black holes break all the supersymmetries, but in some cases they preserve part of the supersymmetries.
 \end{abstract}
\newpage

\section{Introduction}

Symmetries of spacetime play an essential role in Einstein's general relativity. For example, the maximally symmetric spacetime can often be taken as the vacuum of a theory. And the isometries of a black hole allow us to define the conserved quantities of the background and a test particle. Moreover, in a gravity theory with supersymmetry,  supersymmetric black holes have better ultraviolet behaviors and have been one of central subjects in supergravity and string theory. Furthermore, in the context of AdS/CFT correspondence, the supersymmetries and isometries  are the guide lines to find the bubbling geometry.

The notion of symmetry becomes tricky in a theory of higher spin fields. Different from usual Einstein gravity, the gauge transformation of the metric field involves the higher spin fields such that the usual notions of  geometry, such as diffeomorphism and isometries, do not make much sense in the higher spin gravity. One has to find gauge invariant way to define the symmetries of a classical configuration. In the well-known AdS$_4$ Vasiliev's higher spin theory, the fact that the higher spin gauge transformations are quite involved and there is short of classical solutions hinders us from investigating this issue. %even thoug on the field theory side the presence of higher spin current constrains the theory very much\cite{Maldecena:2011,Maldecena:2012}.
Fortunately in AdS$_3$ the higher spin gravity is much better under control in many aspects.

%It could be expected that the study of symmetry in

%It is important to understand the symmetries of the classical solutions in the higher spin gravity theory.

The higher spin gravity in AdS$_3$ has been developing quickly in the past few years. One nice feature of the AdS$_3$ higher spin gravity is that the original Vasiliev theory\footnote{See \cite{Vasi} for a review.} could be cast into a Chern-Simons gravity on a high spin algebra\cite{Blen1,Blen2}, and could even be truncated to a theory  on a finite rank Lie algebra, if not considering the matter scalar field. Therefore the classical solutions without scalar hair can be constructed explicitly. More interestingly it was proposed\cite{Gaberdial} that the Vasiliev theory in AdS$_3$ could be  holographically dual to a 2D ${\cal W}_{N,k}$ minimal model at the boundary.
  Up till now, there are two kinds of limit studied on this duality in the literature. The first one is the  't Hooft limit, which is obtained by taking $N,k\to\infty$ while keeping the `t Hooft coupling $\lambda=\frac{N}{N+k}$ fixed. Under this limit the boundary theory is unitary, but the bulk theory has some troubles in counting the dual light states. See the review \cite{review} for recent developments and references therein. For a recent proposal, see \cite{Chang:2013izp}. The other one is the semi-classical limit\cite{Triality, Perlmutter:2012}, which is obtained by taking $c\to\infty$ while keeping $N$ fixed. In taking this limit, the level $k$  in the boundary CFT has to be negative and there are states with negative conformal weights. Hence, in this case, the theory is non-unitary. However, the bulk theory in the semi-classical limit is simpler than the one in the `t Hooft limit as the gauge group is of finite rank. Hence it allows us to investigate the HS/CFT correspondence in detail. As the first step to check the correspondence, one has to match the spectrum on two sides. On the CFT side, the minimal model has various representations characterized by
$(\Lambda_+;\Lambda_-)$, where $\L_\pm$ are the integrable highest weight representations of the affine algebra $su(N)$ at level $k$ and $k+1$. Among them, the primary states in $(0,\L_-)$ are of particular importance. In the semiclassical limit, these primary states have conformal dimensions proportional to the central charge, indicating their non-perturbative nature. It was proposed in \cite{Perlmutter:2012} that the states $(0,\L_-)$ correspond to  smooth conical defects(surplus) in the bulk AdS$_3$ higher spin gravity. The states $(\L_+,0)$ corresponds to scalar perturbation and the general states $(\L_+,\L_-)$ correspond to bound states of the scalar perturbation and the conical defects.

The smooth conical defects(surplus)\cite{Gopakumar:2012} are classical solutions of the AdS$_3$ higher spin gravity. They have the same topology as the global AdS$_3$, with a contractible spatial circle. As the usual geometric notions break down in the higher spin gravity, one has to use gauge invariant quantities to characterize these solutions. In the case of conical defect, a well-defined quantity is the holonomy of the gauge field along the contractible spatial circle. The smooth conical defect has a trivial spatial holonomy such that the corresponding gauge potential is not singular. As the corresponding states have maximally degenerate null vectors, the smooth conical defects are expected to have maximal higher spin symmetry, as the global AdS$_3$ vacuum. %The meaning of maximal higher spin symmetry will be made precise in section 2.

Another interesting class of classical solutions is higher spin black hole\cite{Per kraus:2011}. Different from the conical defects, the spatial circle of the black hole is not contractible but its time circle is.  The smoothness of the higher spin black hole requires that the holonomy of the gauge field along the time circle is in the center of the gauge group. More interestingly, the trivial thermal holonomy leads to consistent thermodynamics for the higher spin black holes. On the other hand, the spatial holonomy of the gauge field for the higher spin black hole is not trivial. Hence, there is an interesting question: what is the information encoded in the spatial holonomy?

%In the discussion of classical solutions of AdS$_3$ higher spin gravity, one important

%One interesting class of objects in  is the conical defects(surplus)\cite{Gopakumar:2012}.

 %Now we have identified two kinds of classical solutions in higher spin gravity: conical defects(surplus)\cite{Gopakumar:2012} and higher spin black holes\cite{Per kraus:2011}.

%The interesting fact is that these two kinds of solutions are related to the quantity which is called the holonomy. The smooth conical defects (surplus) require the holonomy around the spatial circle to be in the center of the corresponding group, and the smooth black holes require the holonomy around the thermal circle to be in the center of the corresponding group. On the one hand, it is quite natural from the point of gauge invariance. The Chern-Simons higher spin gravity is a topological theory and all the information is encoded in the gauge invariant quantity: holonomy. On the other hand, it is quite puzzle. For example, there are two independent of holonomy, the holonomy around the thermal circle and the holonomy around the spatial circle for a higher spin black hole. To define a  higher spin black hole, it is enough to use the thermal holonomy and then we can find that it is consistent with the first law of thermodynamics. The fact is that we even needn't use the spatial holonomy. However, one can easily shown that the spatial holonomy is nontrivial for an arbitrary higher spin black hole. Hence, there is an interesting question: what is the information which is encoded in the spatial holonomy?

In fact, the spatial holonomy encodes the symmetry of the solution.  Simply speaking, to determine how many symmetries are kept by the solution, we need to solve the following equation in holomorphic sector\footnote{Certainly we need to consider the similar equation in the anti-holomorphic sector as well}
\be
\d_\L A=d\L+[A,\L]=0,\label{gaugetrans}
\ee
where $A$ is the flat connection.
Locally the above equation could always be solved. To have a well-defined gauge transformation, we need to impose periodic boundary condition on the gauge parameter $\L$. In the end we obtain the following relation
\be
e^{-2\pi a_\phi}\l e^{2\pi a_\phi}=\l, \label{holo}
\ee
where $a_\phi$ is the $\phi$-component in $A$ and $\l$ is some constant matrix valued in the Lie algebra of the gauge group.  When the spatial holonomy  is in the center of the corresponding group, the solution is of maximal higher spin symmetries in the sense that $\l$ has maximal number of degrees of freedom. In other words, the smooth conical defect(surplus) is the maximally symmetric solution in the higher spin gravity.
%To answer the question, we need a deeper understanding of the relationship of the conical defects(surplus) and spatial holonomy. The conclusion is that  %This is interesting for two reasons. At first,
%symmetry plays an important role $AdS/CFT$ correspondence. The maximally symmetric solution, such as $AdS_4\times S^7, AdS_5\times S^5, AdS_7\times S^4$, is quite essential for the validity of the duality. Our results show that conical defects(surplus) are actually the higher spin cousin of these maximally symmetric solutions.
Actually, we show how to obtain the smooth conical defects by searching for the maximally symmetric solutions of the corresponding higher spin gravity. This turns out to be a quite effective method. We use this method to find out the smooth conical defects in $sl(N), so(2N+1),sp(2N),so(2N),g_2$ gravity theories, as well as the ones in $sl(N|N-1)$ and $osp(2N+1|2N)$ supergravities. Moreover we establish an one-to-one match between the conical defects and the highest weight representations of dual group in all cases. This exact match of the spectrum suggests that there may exist a correspondence between the finitely truncated higher spin gravities, possibly coupled to scalar matter, with some kinds of minimal models. When the spatial holonomy is non-trivial, as in the case of higher spin black hole, the solution is partially symmetric. In the case of generic higher spin black hole, the constant matrix have to be valued only in Cartan subalgebra of the gauge group, showing the black hole could have well-defined global charges.

In a higher spin supergravity, the spatial holonomy not only encodes the symmetries of the solution but also supersymmetries preserved by the solution.
The supersymmetric configurations are of particular interest in a supersymmetric theory.
They often have nice properties and are easier to deal with. The supersymmetric solutions in the higher spin (super-)gravity have been discussed recently in \cite{Justin:susy,Tan:2012,Hikida:3}. In \cite{Justin:susy}, the higher spin generalization of Killing spinor equation has been proposed. And in \cite{Hikida:3}, the maximally supersymmetric conical defects have been discussed in $sl(N|N-1)$ gravity. In the Chern-Simons supergravity, the gauge group is a supergroup. As a result, the constant matrix $\l$ in (\ref{holo}) should be valued in the supergroup, with its fermionic sector being labeled by $\epsilon$. Taking into account of the boundary condition on the fermionic sector of $\L$, which could be either periodic or anti-periodic, one obtain the following relation on the spinor
 \be
e^{-2\pi a_\phi}\epsilon e^{2\pi a_\phi}=\pm\epsilon, \label{holof}
\ee
From this relation, one may read out how many supersymmetries the solution preserve\cite{Justin:susy}.
 In usual supergravity, the extremal black holes are often supersymmetric. Therefore it is interesting to investigate if the extremal higher spin black holes can keep part of the supersymmetries as well. It turns out to be true, but the story is more interesting. Using brute force, one may solve the generalized Killing spinor equation, which is the fermionic part of (\ref{gaugetrans}), to find the supersymmetric higher spin black holes. These supersymmetric black holes are exactly the ones obtained by solving holonomy equations (\ref{holof}) imposed by spatial holonomy.

%This turns out to be correct for higher spin black holes. For example, the susy information can be found by solving the killing spinor equation by brute force. We find that the susy information can also be found by

The structure of this paper is as follows. In section 2, we clarify the relationship of the spatial holonomy and the maximally symmetric solution, then we search for the maximally symmetric solutions in various higher spin (super-)gravity theories. In section 3, we discuss the partially (super-)symmetric solution. In section 4, we explore the black holes in the higher spin (super-)gravity theories with gauge group $osp(3|2)$, and $sl(3|2)$ and study their supersymmetries. We end this paper by some conclusion and discussions. The appendix collects the convention we use in this paper.

\section{Maximally Symmetric Solutions}

The motivation to study maximally symmetric solutions in a higher spin gravity is two-fold. Firstly, in the Einstein gravity, the maximally symmetric solution always plays important role. It is defined to be the spacetime with maximal number of globally defined Killing vectors. Actually it is unique for fixed signature and dimension, and is often regarded as the vacuum of a theory. In a supergravity theory, it could carry maximal number of Killing spinors and thus could be the maximally supersymmetric configurations. As the higher spin theory is a generalization of conventional gravity theory, it is quite interesting to search for the maximally symmetric solutions in the higher spin gauge theories. Secondly, from the HS$_3$/CFT$_2$ correspondence in the semiclassical limit, the maximally symmetric configurations in the bulk higher spin gravity should correspond to the non-perturbative state $(0,\L_-)$, which has maximally degenerate null vectors. This point has been carefully investigated in a recent paper\cite{Perlmutter:2012}. So searching for the maximally symmetric solutions in the higher spin gravity is a well posed and important problem. In this section, we first study this issue in the bosonic higher spin theory and then turn to the higher spin supergravity.

%For example,Minkowski spacetime is the background of non-gravitational physics. Other important examples in $AdS/CFT$ are $AdS_4\times S^7, AdS_5\times S^5, AdS_7\times S^4$, they preserve the maximal symmetry\footnote{When we include supersymmetry, they preserve maximal supersymmetry, too.}.  Second, from the CFT side, the maximally symmetric solution is always important,too. The symmetry in the bulk gravity maps to null states in the CFT side.  Symmetry is quite important in the $HS/CFT$ correspondence. This comes from the fact that the higher spin symmetry is always infinite dimensional hence becomes quite powerful. In $AdS_4$, the exact higher spin symmetry has been shown to be quite powerful to constraint the correlation function\cite{Maldecena:2011}. In the case with a slightly higher spin symmetry breaking, it is also powerful to constraint the correlation functions\cite{Maldecena:2012}.

\subsection{Maximally Symmetric Solution in Bosonic Higher Spin Gravity}
First of all, we need to define what the maximally symmetric solution is in a bosonic higher spin gravity. In this section, we are searching for the solutions with the topology $D^2\times R$. We will use coordinates $(\rho, \phi, t)$, with $\phi\sim \phi+2\pi$ being a contractible cycle, and $z=x^+=t+\phi, \bar{z}=x^-=t-\phi$. We just focus on the holomorphic part of the solution here, and choose the gauge group $SL(N)$ to illustrate the problem.

 To compare with states in the CFT, the solution should be asymptotic to global $AdS_3$, namely
\be
A-A_{AdS_3}\sim\mathcal{O}(1).\label{asym}
\ee
We can choose the highest weight gauge to set the gauge field to be of the form\cite{Theisen:2011}
    \be
    A=b^{-1}a b +L_0d\rho\label{gauge1}
    \ee
    where $b=\exp{L_0\rho}$ and $a=a_+ dz$ with
    \be
    a_+=L_1+\sum_{s=2}^{N}\mathcal{W}_s W^s_{-s+1}.\label{gauge2}
    \ee
    The definition of $W^s_{-s+1}$ can be found in the appendix. We are interested in the solutions with constant $a$. And it has been shown by the asymptotic symmetry analysis \cite{Theisen:2011, Henneaux:2011} that $\mathcal{W}_s$ can be identified to  the spin $s$ charge. In the following, we do not distinguish $\mathcal{W}_2$ and $\mathcal{L}$ that was used in many other references.

 The solution parameterized by (\ref{gauge1},\ref{gauge2}) has an asymptotic $W_N$ symmetry which is generated by the gauge transformation that preserve the asymptotic $AdS_3$ boundary condition (\ref{asym}). To determine how many higher spin symmetries is kept by the solution, we need to solve the following equation
  %Here the maximally symmetric solution should have maximal higher spin symmetry, that is
    \be
    \delta_{\Lambda}A=d\Lambda+[A,\Lambda]=0,\label{sym}
    \ee
    where $\L$ is the the parameter of the gauge transformation.
    This equation can always be solved locally.
    The $\rho$ component of the equation (\ref{sym}) can be solved by
    \be
    \Lambda=b(\rho)^{-1}\Lambda_0 b(\rho).
    \ee
    The $+,-$ component equation of (\ref{sym}) indicate that
    \be
    \Lambda_0=\exp -(a_+ z)\lambda \exp(a_+z)
    \ee
    where $\lambda$ is a constant matrix taking value in $sl(N)$. Note that $\Lambda$ is nothing but the higher spin generalization of Killing vector in conventional gravity. To be globally defined, it should satisfy the periodic boundary condition in the spatial $\phi$ direction
     \be
    \Lambda(z+2\pi)=\Lambda(z).
    \ee
    This leads to the constraint
    \be
    \exp-(2\pi a_+)\lambda\exp(2\pi a_+)=\lambda.
    \ee
    In order to have a maximally higher spin symmetric solution, $\lambda$ should be an arbitrary constant matrix valued in $sl(N)$. This leads to the requirement that the holonomy of the gauge field along the spatial $\phi$ cycle
    \be
    H_{\phi}(A)=\exp {\oint_{\phi}A}\sim \exp{2\pi a_+}
    \ee
     must be in the center of $SL(N)$.  In other words, if $H_{\phi}(A)$ is trivial, then the solution is maximally symmetric. As a consequence, $a_+$ must be diagonizable and has different eigenvalues.

    The above discussion is  consistent with the results in pure gravity. In the AdS$_3$ Chern-Simons gravity, the gauge transformations of the gauge fields encodes the information of local Lorentz transformation and diffeomorphism. The maximally symmetric solution defined above is exactly the global AdS$_3$, and the constant $SL(2,\mathbb{R})$ actually correspond to  the holographic one in the isometry group $SO(2,2)\simeq SL(2,\mathbb{R}) \times SL(2,\mathbb{R})$.

     Obviously, the above discusion is valid for other gauge groups. Let us discuss them case by case.

  \subsubsection{$SL(N,\mathbb{R})$ and $SL(N,\mathbb{C})$}

     The center of $SL(N,\mathbb{R})$ is different for odd or even $N$. For odd $N$, its center is $I$, so we have
        \be
        H_{\phi}(A)\sim \exp{2\pi a_+}\sim I.\label{holo1}
        \ee
        The equation (\ref{holo1}) only depends on the eigenvalues of $a_+$. Assumed the eigenvalue of $a_+$ to be
        \be
        a_+\sim diag(\th_1,\th_2,\cdots,\th_N),
        \ee
        the holonomy condition (\ref{holo1}) tells us
        \be
        \th_1=i n_1,\ \th_2=i n_2,\cdots \th_N=i n_N\label{re}
        \ee
        with $n_1,n_2,\cdots,n_N\in \mathbb{Z}$. For $a_+$ to be  diagonalizable, we require $n_i\neq n_j$ for $i\neq j$.
        It is also convenient to assume $n_1>n_2\cdots>n_N$.  %The condition (\ref{re}) is the consequences of the holonomy condition (\ref{holo1}). This automatically preserves maximal symmetry of the theory.
        Note that the traceless condition of $sl(N)$ requires
        \be
        \sum_{i=1}^{N}n_i=0\label{re2}.
        \ee
        Moreover, for the $SL(N,\mathbb{R})$ case, $a_+$ must be real, which impose further conditions on $n_i$.
        Let us find the consequences of (\ref{re})(\ref{re2}) for the higher spin charges $\mathcal{W}$. We take $N=3$ to illustrate the point. In this case, (\ref{gauge2}) becomes
        \be
        a_+=L_1+\mathcal{L}L_{-1}+\mathcal{W}_3W_{-2}^3.
        \ee
        Note $\mathcal{L},\mathcal{W}_3$ are proportional to the trace of the power of $a_+$.
        Since we require the charge $\mathcal{W}_3$ to be real, the $n_i$ should be
        \be
        (n_1,n_2,n_3)=(n, 0, -n).
        \ee
        Thus we have
        \be
        \mathcal{L}=\frac{n^2}{4},\ \mathcal{W}_3=0.
        \ee
         Here $n=1$ corresponds to  global $AdS_3$ embedded in $SL(3,\mathbb{R})$, and the other solutions with $n\ge2$ correspond to smooth conical surplus studied in \cite{Gopakumar:2012}. The vanishing of spin 3 charge originates from the reality condition on the gauge potential. More generally for all odd $N$, the condition of a real connection always leads to vanishing odd spin charges.

        For even $N$, its center is $\pm I$, so we have
        \be
        H_{\phi}(A)\sim \exp{2\pi a_+}\sim \pm I.\label{holo2}
        \ee
As before, the condition (\ref{holo2}) is relevant to the eigenvalue of $a_+$. We assume
\be
a_+\sim diag(\th_1, \th_2,\cdots,\th_N).
\ee
If the holonomy is $I$, then
\be
\th_i=i n_i,\  n_i\in\mathbb{Z}.
\ee
If the holonomy is $-I$, then
\be
\th_i=i (n_i+\frac{1}{2}),\ n_i\in\mathbb{Z}.
\ee
%The traceless  condition is still $\sum_{i=1}^N\th_i=0$.
In the case of $N=2$, we  find that
\be
\mathcal{L}=\frac{n^2}{4},\ n\in\mathbb{Z}^+.
\ee
The holonomy is $-I$ for odd $n$, and $I$ for even $n$. When $n=1$, the solution is just global AdS$_3$, while when $n\ge 2$, the solutions are the smooth conical surplus.

We note that all the maximally symmetric solutions in $N=2,3$ have vanishing higher spin charge. However, this situation changes when $N\ge4$. We take $N=4$ as an example. We find the eigenvalues of $a_+$ to be
\be
(\th_1,\th_2,\th_3,\th_4)=i(n_1,n_2,-n_2,-n_1),
\ee
where $n_i$ can be intergar or half intergar, depending on the holonomy, or the choice of the center. The spin 3 charge $\mathcal{W}_3$ is still zero, but the even spin 2 and spin 4 charges are nonzero
\bea
\mathcal{L}&=&\frac{1}{10}(n_1^2+n_2^2)\propto C_2(n)\nn\\
\mathcal{W}_4&=&\frac{82n_1^2 n_2^2-9n_1^4-9n_2^4}{3600}\propto (C_4(n)-\frac{C_4(\rho)}{C_2(\rho)^2}C_2(n)^2)
\eea
where the $C_2, C_4$ are the Casimirs in $SL(4,\mathbb{R})$ and $\rho$ is the corresponding Weyl vector . These are the conical defects or surplus in the spin 4 gravity studied in \cite{Gopakumar:2012}. They have nonvanishing higher spin charges. The same feature holds for all the maximally symmetric solutions when $N\ge4$.

Note that our maximally symmetric solutions in the higher spin gravity are just the smooth conical defects(surpluses) studied in \cite{Gopakumar:2012,Perlmutter:2012}.  However, the work in \cite{Gopakumar:2012} was motivated in matching the conical defects to the primary states in CFT with the same global charges, while here we have shown that the smooth conical defects(surpluses) should have been discovered by simply symmetry consideration.

 To match the primary states in the $CFT$ side, the $SL(N,\mathbb{C})$ case is also important. One need do Euclidean continuation to match the spectrum. The center of $SL(N,\mathbb{C})$ is $e^{\frac{-2\pi i m}{N}}1_{N\times N}$, hence the eigenvalues of $a_+$ are
    \be
    \th_i=i(m_i-\frac{m}{N}),\ m_i\in\mathbb{Z}, i=1,2,\cdots, N.
    \ee
    The traceless condition of $SL(N,\mathbb{C})$ leads to
    \be
    \sum_{i=1}^{N}m_i=m.
    \ee
    The $m_i$ can be shifted to set $m_N=0$. To be in match with the CFT states, one needs the Young diagram of $su(N)$. A Young diagram of $su(N)$ includes $N-1$ rows, each row has $r_i(r_N=0)$ boxes. This Young diagram is in one to one correspondence with the highest weight state $(0,\Lambda_-)$ with
    \be
    \Lambda_-=(\Lambda_1,\cdots,\Lambda_N)
    \ee
    where
    \be
    \Lambda_i=r_i-\frac{\sum_{i=1}^{N}r_i}{N}.
     \ee
     To relate it to the gravity solutions, we can define
    \be
    r_i=m_i-(N-i)
    \ee
    such that the eigenvalues of the holonomy could be rewritten as
    \be
    -i\th_i=r_i-\frac{\sum_{i=1}^{N}r_i}{N}+\frac{N+1}{2}-i=\Lambda_i+\rho_i
    \ee
    where $\rho_i=\frac{N+1}{2}-i$ is the Weyl vector of $su(N)$. Hence we find a one to one correspondence between the bulk maximally symmetric solution and the highest weight state $(0,\Lambda_-)$.

\subsubsection{$Sp(2N,\mathbb{R})$ and $Sp(2N,\mathbb{C})$}

 These cases are motivated by the proposed even spin minimal model hologaphy\cite{Ah,Gopa,Gaberdial:2012}. Note that the center of $Sp(2N,\mathbb{R})$ and $Sp(2N,\mathbb{C})$ are the same, which can be $\pm I$. Then the holonomy now is
    \be
    H_{\phi}(A)\sim \exp(2\pi a_+)\sim \pm I
    \ee
    The eigenvalues of $a_+$ can be parameterized as
    \be
    a_+\sim diag(\th_1,\th_2,\cdots,\th_N,-\th_N,-\th_{N-1},\cdots,-\th_1).
    \ee
    If the holonomy is chosen to be $I$, then
    \be
    \th_i=i n_i, \ n_i\in\mathbb{Z}.
    \ee
    If the holonomy is chosen to be $-I$, then
    \be
    \th_i=i(n_i+\frac{1}{2}), \ n_i\in\mathbb{Z}.
    \ee
     To make sure $a_+$ is diagonalizable, one has an additional requirement that $n_i\neq n_j$ for all $i,j$. Hence it is convenient to assume $n_1>n_2>\cdots >n_N$.

%It would be very interesting to match the solutions found above to the CFT highest weight states. We propose that the CFT dual to $sp(2N)$ gravity is a coset model
%\be
%\frac{SO(2N+1)_k\oplus SO(2N+1)_1}{SO(2N+1)_{k+1}}.
%\ee
On the other hand,  a representation of $so(2N+1)$ can be parametrized by its highest weight, as $N$ numbers $r_1\geq r_2\geq\cdots\geq r_N\geq 0$  (see 1.65 in \cite{Dictionary}). There are two kinds of representations: the vector representation with all $r_i$'s being integers and the spinor representation with all $r_i$'s being half-integer. The relation between the weight $\Lambda$ and $r_i$ turns out to be:
    \be
    \Lambda=\sum_{i=1}^{N-1}(r_i-r_{i+1})\lambda_i+2 r_N\lambda_N
    \ee
where $\lambda_i$ is the $i$-th fundamental weight:
\bea
\lambda_1&=&e_1\nn\\
\lambda_2&=&e_1+e_2\nn\\
&\vdots&\nn\\
\lambda_{N-1}&=&e_1+e_2+\cdots+e_{N-1}\nn\\
\lambda_N&=&\frac{e_1+\cdots+e_N}{2}
\eea

The Weyl vector is
\be
\rho=\sum_{i=1}^N \lambda_i=\sum_{i=1}^N (N+\frac{1}{2}-i)e_i.
\ee
Hence in this case the correspondence reads
\be
-i\theta_i=\Lambda_i+\rho_i=r_i+N+\frac{1}{2}-i.
\ee
Note that the vector representations of $so(2N+1)$ exactly correspond to half-integer valued $-i\th_i$'s, which are smooth conical defects whose holonomies are in the  center $-I$ of $Sp(2N,\mathbb{R})$. For example, the trivial representation $r_1=r_2=\cdots=r_N=0$ corresponds to the $AdS_3$ vacuum with $a_+=i\hspace{1mm}diag(N-\frac{1}{2},N-\frac{3}{2},\cdots,\frac{1}{2},-\frac{1}{2},\cdots,-(N-\frac{1}{2}))$. On the other hand, the spinor representations of $so(2N+1)$ have half-integer valued $r_i$, hence exactly correspond to the conical defects whose holonomy is in the center $I$ of $Sp(2N,\mathbb{R})$. Therefore we see that  each highest weight state of $so(2N+1)$ is in exact match with the smooth conical defects in the higher spin gravity with gauge group $Sp(2n,\mathbb{R})$ or $Sp(2n,\mathbb{C})$.

\subsubsection{$SO(2N+1,\mathbb{R})$ and $SO(2N+1,\mathbb{C})$}.

This is another realization of even spin gravity. Again, the centers of $SO(2N+1,\mathbb{R})$ and $SO(2N+1,\mathbb{C})$ are the same, being $I$.  The eigenvalues of $a_+$ can be parameterized by
    \be
    a_+\sim diag(\th_1,\cdots,\th_N,0,-\th_N,\cdots,-\th_1).
    \ee
    The holonomy  requires
    \be
    \th_i=i n_i, \ n_i\in\mathbb{Z}.
    \ee
As before, the diagonalizable condition of $a_+$ requires $n_i$ are all distinct numbers: $n_1>n_2>\cdots>n_N$.

%In this case, we propose the dual CFT being a coset
%   \be
%\frac{Sp(2N)_k\oplus Sp(2N+1)_1}{Sp(2N+1)_{k+1}}.
%\ee
   The highest weights of $sp(2N)$ representations are parametrized by $N$ integers $r_1\geq r_2\geq\cdots\geq r_N\geq 0$  (see 1.66 in \cite{Dictionary}). The relation between the weight $\Lambda$ and $r_i$ is:
    \be
    \Lambda=\sum_{s=1}^{N-1}(r_s-r_{s+1})\lambda_s+r_N\lambda_N=\sum_{i=1}^{N}r_i e_i,\nn\\
    \ee
where the fundamental weights $\lambda_i$ are:
\bea
\lambda_1&=&e_1,\nn\\
\lambda_2&=&e_1+e_2,\nn\\
&\vdots&\nn\\
\lambda_{N}&=&e_1+e_2+\cdots+e_N.
\eea
The Weyl vector is
\be
\rho=\sum_{i=1}^N \lambda_i=\sum_{i=1}^N (N+1-i)e_i\nn.
\ee
In this case the correspondence reads
\be
n_i=-i\theta_i=\Lambda_i+\rho_i=r_i+N+1-i.
\ee
Therefore we find an exact match of the smooth conical defects in $SO(2N+1)$ Chern-Simons gravity and the highest weight representations of $sp(2N)$.

Note there is an interesting ``duality'' between $B_N$ and $C_N$ Lie algebras: the smooth conical defects in $B_N$ gravity could correspond to a $C_N$-type highest weight representation and the smooth conical defects in $C_N$ gravity could correspond to a $B_N$-type highest weight representation. % in finite $N$ and a classical limit with $c\rightarrow\infty$.
%This novel duality does not contradict with the existing even spin holography. Actually in \cite{Gaberdial:2012} it has been shown that the coset $\frac{sp(2N)_k\oplus sp(2N)_{-1}}{sp(2N)_{k-1}}$ and $\frac{so(2N+1)_k\oplus so(2N+1)_1}{so(2N+1)_{k+1}}$ both possess $\mathcal{W}^e_\infty$ symmetry in the 't Hooft limit.

\subsubsection{$SO(2N,\mathbb{R})$ and $SO(2N,\mathbb{C})$}

%This case has been discussed in \cite{Ah,Gopa}. It has been proposed that the higher spin gravity on $so(2N)$ is dual to the coset
%\be
%\frac{SO(2N)_k\oplus SO(2N)_1}{SO(2N)_{k+1}}.
%\ee

The groups  $SO(2N,\mathbb{R})$ and $SO(2N,\mathbb{C})$ have the same center $\pm I$. For the smooth conical defects, $e^{2\pi a_+}$ has to be in the center. The generic diagonalized form of $a_+$ can be written as:
\be
a_+=diag(\th_1,\th_2,\cdots,\th_N,-\th_N,\cdots,-\th_1)=diag(in_1,in_2,\cdots,in_N,-in_N,\cdots,-in_1).
\ee
 If the spatial holonomy being in the center $I$, $n_i$'s take value in $\mathbb{Z}$, while if the spatial holonomy being the center $-I$, $n_i$'s take value in $\mathbb{Z}+\frac{1}{2}$. The diagonalizable condition requires $n_1>n_2> \cdots >n_N\geq 0$.

The fundamental weights in $so(2N)$ are:
\bea
\lambda_1&=&e_1\nn\\
\lambda_2&=&e_1+e_2\nn\\
&\vdots&\nn\\
\lambda_{N-2}&=&e_1+e_2+\cdots+e_{N-2}\nn\\
\lambda_{N-1}&=&\frac{1}{2}e_1+\frac{1}{2}e_2+\cdots+\frac{1}{2}e_{N-1}-\frac{1}{2}e_N\nn\\
\lambda_N&=&\frac{1}{2}e_1+\frac{1}{2}e_2+\cdots+\frac{1}{2}e_{N-1}+\frac{1}{2}e_N.
\eea
And the Weyl vector is
\be
\rho=\sum_{i=1}^N \lambda_i=\sum_{i=1}^N (N-i)e_i.
\ee

 A highest weight representation of $so(2N)$ is labelled by $N$ numbers $r_1\geq r_2\geq \cdots\geq r_N\geq 0$, where $r_i$ could all be integer or half-integer\cite{Gopa} %there are actually two kinds of irreducible representations: self-dual one $[S(r)]$ and anti-self-dual one $[A(r)]$\cite{ma2007group}. For the self-dual one $[S(r)]$,
\bea
&&\Lambda=\sum_{i=1}^{N-2}(r_i-r_{i+1})\lambda_i+(r_{N-1}-r_N)\lambda_{N-1}+(r_{N-1}+r_N)\lambda_N=
\sum_{i=1}^{N}r_i e_i\nn\\
&&(\Lambda+\rho)_i=r_i+N-i
\eea
%And for the anti-self-dual one $[A(r)]$,
%\bea
%\Lambda&=&\sum_{i=1}^{N-2}(r_i-r_{i+1})\lambda_i+(r_{N-1}+r_N)\lambda_{N-1}+(r_{N-1}-r_N)\lambda_N\nn\\
%&=&\sum_{i=1}^{N-1}r_i e_i-r_N e_N\nn\\
%(\Lambda+\rho)_i&=&\left\{\begin{array}{cl}r_i+N-i&i\leq N-1\\-r_N&i=N\end{array}\right.
%\eea
When all $r_i$'s are integers, they correspond to vector representations and
if they are all half-integers, they correspond to spinor representations.

There is an one-to-one correspondence between $n_i$ and $r_i$%. Note that a ``-'' sign on $r_N$ does not change $(\Lambda+\rho)^2$ hence $[S(r)]$ and $[A(r)]$ for the same n-tuple $r$ correspond to a single conical defect on the gravity side. Despite of this peculiar phenomenon, it is straightforward to see
 \be
 n_i=-i\th_i=(\Lambda+\rho)_i=r_i+N-i,
  \ee
  both of which have same range. For example, the vacuum configuration has the eigenvalues $(n_1,n_2,n_3,n_4)=(3,2,1,0)$, which exactly corresponds to the trivial representation $(r_1,r_2,r_3,r_4)=(0,0,0,0)$. Therefore we establish the correspondence between the highest weight representations of $so(2N)$ and the smooth
  conical defects in $SO(2N)$ higher spin gravity.

%This suggests a correspondence between $so(2N)$ gravity and a $\frac{so(2N)\oplus so(2N)}{so(2N)}$ coset model.

\subsubsection{$G_2(\mathbb{R})$ and $G_2(\mathbb{C})$}

 In this case, the corresponding  higher spin gravity has only spin 2 and spin 6 fields\cite{truncated}. The centers of $G_2(\mathbb{R})$ and $G_2(\mathbb{C})$ are both trivial, hence  the diagonalized form of $a_+$ is
 \be
a_+=diag(in_1,in_2,in_3,0,-in_3,-in_2,-in_1).
\ee
 and $\theta_i=i n_i$ with $n_i\in \mathbb{Z}$. In this case the eigenvalue equation of $(L_1+\mathcal{L}L_{-1}+\mathcal{W}_6 W^6_{-5})$ is:
    \be
    \lambda[(\lambda^2+4\mathcal{L})(\lambda^2+16\mathcal{L})(\lambda^2+36\mathcal{L})+172800\mathcal{W}_6]=0,
    \ee
whose roots are $0,\pm i n_1,\pm i n_2, \pm i n_3$. If the spin-6 charge $\mathcal{W}_6=0$ then clearly one requires $\mathcal{L}=n^2/4$ and $n_1=3n,n_2=2n,n_3=n$. If the spin-6 charge $\mathcal{W}_6$ is non-vanishing, the solutions need more efforts. From the algebraic relations between $n_i$ and $\mathcal{L}$
    \bea
    &&n_1^2+n_2^2+n_3^2=56\mathcal{L}\nn\\
    &&n_1^2 n_2^2+n_2^2 n_3^2+n_3^2 n_1^2=784\mathcal{L}^2,
    \eea
    we find $n_1^4+n_2^4+n_3^4=2(n_1^2 n_2^2+n_2^2 n_3^2+n_3^2 n_1^2)$, which requires that  one of $n_i$'s equals to the sum of the other twos. Without losing generality, we choose $n_1=n_2+n_3$ and let $n_1>n_2>n_3>0$. The $n_i\neq n_j$ requirement also comes from the diagonalizable condition. Hence the maximally symmetric backgrounds are parametrized by two positive integers $n_2$ and $n_3$. Accordingly the values of $\mathcal{L}$ and $\mathcal{W}_6$ are respectively:
    \bea
    \mathcal{L}&=&\frac{n_2^2+n_2 n_3+n_3^2}{28}, \nn\\
    \mathcal{W}_6&=&\frac{1}{172800}(n_2^2 n_3^2(n_2+n_3)^2-\frac{36}{343}(n_2^2+n_2 n_3+n_3^2)^3).
    \eea

On the other hand, the representation of $g_2$ is characterized by the highest weight (see 1.63 of \cite{Dictionary}) $\Lambda=\frac{r_1-2r_2}{3}\lambda_1+r_2 \lambda_2$, where the fundamental weights are:
    \be
    \lambda_1=-e_1-e_2+2e_3,\ \ \lambda_2=e_3-e_2.
    \ee
    Therefore we have
    \be
    \Lambda+\rho=(\frac{2}{3}r_2-\frac{1}{3}r_1-1)e_1+(-\frac{1}{3}r_2-\frac{1}{3}r_1-2)e_2+(\frac{2}{3}r_1-\frac{1}{3}r_2+3)e_3.
    \ee
    From the representation theory of $g_2$ it is required
     \bea
     &&r_1+r_2=3n,\ \ r_1,r_2,n\in\mathbb{N}\nn\\
     &&r_1\geq 2r_2.\label{g2constraint}
     \eea
    We make the following identification between $n_i$'s  and $r_i$'s:
    \bea
   n_3&=&-(\frac{2}{3}r_2-\frac{1}{3}r_1-1)\nn\\
   n_2&=&-(-\frac{1}{3}r_2-\frac{1}{3}r_1-2)\nn\\
   n_1&=&(\frac{2}{3}r_1-\frac{1}{3}r_2+3).
    \eea
    Note that $n_1=n_2+n_3$ is ensured. And from these expressions we get
    $r_1=2n_2+n_3-5,r_2=n_2-n_3-1$, satisfying $3|(r_1+r_2)$ automatically. And $r_1-2r_2=3n_3-3$, which is non-negative as long as $n_3\geq 1$.  Therefore we find the pair $(r_1,r_2)$ on the CFT side exactly corresponds to $(n_2,n_3)$ on the gravity side. For example,
    \begin{enumerate}\item The smallest values of $(n_2,n_3)$: $(2,1)$ corresponds to $(r_1,r_2)=(0,0)$, which is the trivial representation of $G_2$. This means the trivial representation corresponds to the global AdS$_3$ vacuum.

    \item $(n_2, n_3)=(3,1)$ corresponds to $(r_1,r_2)=(2,1)$, which is the 7-dimensional representation of $G_2$.

     \item $(n_2, n_3)=(3,2)$ corresponds to $(r_1,r_2)=(3,0)$, which is the 14-dimensional representation of $G_2$.
     \end{enumerate}
There is an exact match of the maximally higher spin symmetric solutions and the highest weight states of $G_2$ representation. %in this case suggests a possible duality between the $G_2$ gravity and a coset model $\frac{g_2\oplus g_2}{g_2}$.

The discussion on $g_2$ case shows that there need special care in dealing with the exceptional Lie group. In principle it is possible to deal with the $F_4, E_6, E_7$ and $E_8$ groups. We do not include them here.
%\end{enumerate}
%\end{enumerate}

\subsection{Maximally Symmetric Solution in Higher Spin Supergravity}

This subsection is to search for maximally symmetric solutions in higher spin supergravity. The asymptotic analysis of the higher spin supergravity has been given in \cite{Cheng, Ray}. The $\mathcal{N}=2$ supersymmetric extension of the HS/CFT duality has been proposed in \cite{Hikida:1}, which relates the three-dimensional $\mathcal{N}=2$ supersymmetric higher spin theory \cite{Pro} to the Kazama-Suzuki minimal model \cite{KS:1,KS:2}. The $\mathcal{N}=1$ version of duality was proposed in \cite{Hikida:2}. Aspects of conical defects and the higher spin black holes have been partly studied in \cite{Justin:susy, Tan:2012, Hikida:3}. Here we would like to search for the maximally symmetric solutions, which are asymptotic to AdS$_3$ and preserve the maximal symmetry of the theory. Without losing generality, we take $sl(N|N-1)$ to give an illustration\footnote{The maximal supersymmetric conical defects in the higher spin $sl(N|N-1)$ gravity have been studied in \cite{Hikida:3}, but our discussions can be extended to other supergroups. For completeness, we include this case.}. Similar to the bosonic case, we need to impose appropriate asymptotic condition on the solution. This condition is the same as (\ref{asym})
\be
A-A_{AdS_3}\sim\mathcal{O}(1)\label{asymp}
\ee
with
\be
A=b^{-1}a b+L_0d\rho.
\ee
But now $a=a_+dz$ is changed slightly as the bosonic spectrum changes,
\be
a_+=L_1+\sum_{s=2}^{N}\mathcal{W}^{(1)}_sW^{(1)s}_{-s+1}+\sum_{s=2}^{N-1}\mathcal{W}^{(2)}_sW^{(2)s}_{-s+1}+\mathcal{V}J
\ee
where $\mathcal{W}^{(1)}_s,\mathcal{W}^{(2)}_s$ are the corresponding higher spin $s$ charges. $\mathcal{V}$ is the $U(1)$ charges. The matrix generators are given in Appendix. Note that we have turned off the fermionic generators as in supergravity when searching for classical solutions.

Similarly, from the requirement
\be
\delta_{\Lambda} A=d\Lambda+[A,\Lambda]=0,
\ee
the gauge parameter could be locally written as
\be
\Lambda=b^{-1}e^{-a_+z}\lambda e^{a_+z}b
\ee
with $\lambda$ is a constant supermatrix taking value in $sl(N|N-1)$.
If we require $\l$ to be arbitrary supermatrix, then the corresponding solution is maximally symmetric.

The gauge parameter $\lambda$ can be decomposed into the bosonic parameter $\xi$ and the fermionic parameter $\epsilon$
 \be
 \lambda=\xi+\epsilon.
 \ee
 Since the background solution $A$ contains only the bosonic generator, the requirements on two parameters $\xi$ and $\epsilon$ are decoupled so can be studied separately. The bosonic gauge parameter should satisfy periodic boundary condition in the spatial $\phi$ direction, while the fermionic gauge parameter should satisfy anti-periodic or periodic boundary condition in the spatial $\phi$ direction,
 \be
 \Lambda_{\xi}(z+2\pi)=\Lambda_{\xi}(z),\ \ \Lambda_{\epsilon}(z+2\pi)=\pm\Lambda_{\epsilon}(z)
 \ee
 or equivalently,
 \be
 e^{-2\pi a_+}\xi e^{2\pi a_+}=\xi,\ e^{-2\pi a_+}\epsilon e^{2\pi a_+}=\pm\epsilon\label{hol}
 \ee
 The bosonic relation tells us that the holonomy
 \be
 H_{\phi}(A)\sim e^{2\pi a_+}
 \ee
 should be the center of the bosonic subalgebra of the corresponding superalgebra. For $sl(N|N-1)$ we are discussing, it is just the center of $sl(N)\oplus sl(N-1)\oplus u(1)$. The fermionic part of (\ref{hol}) is a bit more complex due to the anti-periodic boundary condition. It constrains the holonomy along the $\phi$ cycle as well. For $sl(N|N-1)$, the bosonic supermatrix can be constructed by the anticommutator of two fermionic supermatrix, hence the first condition in (\ref{hol}) will be satisfied automatically provided that the second condition is satisfied. Namely, the maximally supersymmetric solution automatically has maximal bosonic symmetries.

Here we construct the maximally symmetric solutions in two higher spin supergravity to illustrate the previous discussion. We constrain ourselves to real connection. The Euclidean continuation could be important  but we do not include here.
\begin{enumerate}
\item $sl(N|N-1)$. This has been constructed in \cite{Hikida:3}. The same fermionic relation has been obtained, but starting from generalized Killing spinor equation. We will not repeat the details here.
    It turns out the the maximally supersymmetric conical defects are in exact match with the chiral primaries in $N=(2,2) CP^N$ Kazama-Suzuki model.
\item $osp(2N+1|2N)$. The bosonic part is $so(2N+1)\oplus sp(2N)$. The bosonic spectrum includes two copies of spin $2,4,\cdots,2N$. The connection $a_+$ could be diagonalized to be
    \begin{align}
    a_+\sim \left(\begin{array}{cc}
a_{(2N+1)\times(2N+1)}&0\\0&a_{2N\times 2N}
\end{array}\right)
    \end{align}
    where $a_{(2N+1)\times(2N+1)}$ and $a_{2N\times 2N}$ are
    \bea
    a_{(2N+1)\times(2N+1)}&\sim& diag(\th_1,\th_2,\cdots,\th_N,0,-\th_N,\cdots,-\th_2,-\th_1),\\
    a_{2N\times 2N}&\sim& diag(\varphi_1,\varphi_2,\cdots,\varphi_N,-\varphi_N,\cdots,-\varphi_2,-\varphi_1).
    \eea
    From the discussion of the bosonic case, we can include the following two cases.
    \begin{enumerate}
    \item We choose the center to be $1_{so(2N+1)}\times 1_{sp(2N)}$. This can be satisfied by
    \bea
    \th_i&=&i n_i,\ n_i\in\mathbb{Z},\ i=1,\cdots,N\\
    \varphi_i&=&i m_i,\ m_i\in\mathbb{Z},\ i=1,\cdots,N
    \eea
    Then the fermionic boundary condition is periodic.
    \item We can also choose the center to be $1_{so(2N+1)}\times (-1)_{sp(2N)}$. This can be satisfied by
    \bea
    \th_i&=&i n_i,\ n_i\in\mathbb{Z},\ i=1,\cdots,N,\\
    \varphi_i&=&i (m_i+\frac{1}{2}),\ m_i\in\mathbb{Z},\ i=1,\cdots,N.
    \eea
    Then the fermionic boundary condition is anti-periodic.
    \end{enumerate}
    In \cite{Hikida:2}, it was proposed that in the 't Hooft limit the $osp(2N+1|2N)$ high spin supergravity is dual to the $\mathcal{N}=(1,1)$ super coset
    \be
    \frac{so(2N+1)_k\oplus so(2N)_1}{so(2N)_{k+1}}.\label{11coset}
    \ee
    The primary states in this coset are characterized by $(\Lambda,\Xi)$, where $\Lambda$ and $\Xi$ are the highest weight representations of $so(2n+1)$ and $so(2n)$ respectively\cite{Hikida:2}. They are not in match with the
     maximally supersymmetric conical defects found above. Actually the fact that the bosonic sector of $osp(2N+1|2N)$ high spin supergravity involves both $so(2N+1)$ and $sp(2N)$ group suggests that the possible CFT dual could be nontrivial.  
     
   %   should correspond to the
   % chiral primaries in this super coset. However, it is not obvious the primaries and the conical defects are in match in this case. We need further study on this $osp(2N+1|2N)$ high spin supergravity and its dual conformal field theory.
\end{enumerate}
%\end{enumerate}

\section{Partially Symmetric Solution}

In the previous section, we have explored the maximally symmetric solutions in the higher spin gravity with or without supersymmetry, and found that the maximally symmetric solution were exactly  the smooth conical defects which were investigated before. However, as in conventional gravity, not all the allowed solutions preserve maximal symmetries. For example, the black holes in supergravity always have less symmetries than the vacuum solution which is maximally symmetric and generically breaks the supersymmetry completely. In some cases, the extremal black holes may preserve part of supersymmetry. In other words, they are partially symmetric solutions. Here we extend the concept of partially symmetric solution to the higher spin gravity, both in the bosonic and supersymmetric case.

\subsection{Partially Symmetric Solution in Bosonic Higher Spin Gravity}

Here we still choose $SL(N,\mathbb{R})$ as the prototypic model, but the discussion can be easily generalized. Since we would like to include the higher spin black holes, we do not require the solution to be (\ref{gauge2}), but we still require the solution to be of constant $a$. In the gauge (\ref{gauge1}), the configurations we are interested in could be of the form
\be
a=a_+ dz+a_-d\bar{z}\label{ansatz}
\ee
where $a_+$ is the same as (\ref{gauge1}), but a nonvanishing $a_-$ term has been turned on to refer to the higher spin black holes. As the solutions satisfy the flatness condition, $a_+, a_-$ should commute with each other,
\be
[a_+,a_-]=0.\label{eqn}
\ee
Note that we still need to search for the solution of the equation
\be
\delta_{\Lambda}A=d\Lambda+[A,\Lambda]=0.
\ee
The $\rho$-component of the equation still gives us $\Lambda=b^{-1}\Lambda_0b$, while the $+,-$-components of the equation lead to
\be
\Lambda_0=e^{-(a_+z+a_-\bar{z})}\lambda e^{(a_+z+a_-\bar{z})},\label{solution}
\ee
where $\lambda$ is a constant matrix  taking value in $sl(N,\mathbb{R})$. To derive (\ref{solution}) we have used the equation of motion (\ref{eqn}). For $\Lambda$ being well-defined globally, we require the periodic condition
\be
\Lambda(\phi+2\pi)=\Lambda(\phi).\label{perio}
\ee
After some elementary algebra, we find that
\be
e^{-2\pi a_{\phi}}\lambda e^{2\pi a_{\phi}}=\lambda. \label{cond}
\ee
The exponential $e^{2\pi a_{\phi}}$ is actually the holonomy of the gauge potential along spatial circle for our ansatz (\ref{ansatz}).
Up till now, the treatment is the same as before, but this time we do not require $\lambda$ to be arbitrary. If (\ref{perio}) is satisfied for some constant matrix $\lambda$ valued in $sl(N,\mathbb{R})$,  there is no need to require the holonomy $H$ to be in the center of $SL(N,\mathbb{R})$. Obviously, this will lead to the solution that has only partial symmetry\footnote{We should mention that all the solution has maximal symmetry locally, but not all of them preserve the maximal symmetry globally. The globally well-defined symmetry is capture by (\ref{perio}). Hence the precise meaning of partial symmetry we discussed are globally partial symmetry.}.

As the holonomy of the gauge field needs not to be in the center, the matrix $a_\phi$ may not be diagonalizable. The discussion below separates into two cases. If the matrix $a_{\phi}$ can be diagonalized, we assume that its eigenvalues are $(\th_1,\cdots \th_N)$, i.e.
\be
a_{\phi}\sim diag(\th_1,\cdots,\th_N)
\ee
and the eigenvalues differ from each other, namely, $\th_i\not=\th_j$ for $i\not=j$. Using the identity
\be
\exp\big(-2\pi\sum_{k=1}^{N}\th_k E_{kk}\big) E_{ij} \exp\big(2\pi\sum_{k=1}^{N}\th_k E_{kk}\big)=e^{-2\pi(\th_i-\th_j)}E_{ij}
\ee
where $E_{ij}$ is the $N\times N$ matrix that is 1 in the i-th row and j-th colum, otherwise it is 0, then we find  that if
\be
\th_i-\th_j=i n,\ n\in\mathbb{Z}\label{condition}
\ee
for some pairs $(i,j)$.
We can have globally well-defined $\Lambda$, though it may not be arbitrary. For the most general solution, the condition (\ref{cond}) can only be satisfied by $i=j$ with $n=0$. In other words, only diagonal matrix may satisfy (\ref{cond}). Taking into account of the traceless condition, there are only $N-1$ independent solutions. Recall that the Cartan subagebra could be written as the traceless diagonal matrix, the general solution to (\ref{cond}) could be the linear combination of the Cartan generators. This reflects the fact that there are
$N-1$ well-defined global charges in the theory, corresponding to the spin 2, $\cdots$, spin $N$ charges.

If the matrix $a_{\phi}$ cannot be diagonalized, we need to solve the equation (\ref{cond}) from scratch. Note that if
\be
[a_{\phi},\lambda]=0\label{comu}
\ee
then the equation (\ref{cond}) can be satisfied automatically. The equation (\ref{comu}) can always be solved by the traceless function $f(a_{\phi})$ with
\be
\lambda=f(a_{\phi}).\label{fun}
\ee
Since $f$ can always be expanded as
\be
f(a_{\phi})=\sum_{i=2}^{N}c_i (a_{\phi}^i-\frac{1}{N}tr a_{\phi}^i)
\ee
there are always  $N-1$ independent solutions which are captured by the constants $c_i$. We emphasize that there may be symmetry enhancement for special configuration.

\subsection{Partially Symmetric Solution in Higher Spin Supergravity}

The discussion in the previous subsection is a warm-up to the more interesting case we will consider now. We use the superalgebra $sl(N|N-1)$ as our prototypic model. The solution of $\Lambda_0$ is the same as (\ref{cond}), but $\lambda$ can be decomposed into the bosonic part $\xi$ and the fermionic part $\epsilon$,
\be
\lambda=\xi+\epsilon
\ee
The boundary condition is now
\be
\Lambda_{\xi}(\phi+2\pi)=\Lambda_{\xi}(\phi),\ \Lambda_{\epsilon}(\phi+2\pi)=\pm\Lambda_{\epsilon}(\phi).
\ee
The discussion of the bosonic part $\Lambda_{\xi}$ goes through parallel to the previous subsection, so we only focus on the fermionic part $\Lambda_{\epsilon}$. The periodic or anti-periodic boundary condition leads to
\be
e^{-2\pi a_{\phi}}\epsilon\  e^{2\pi a_{\phi}}=\pm\epsilon. \label{condf}
\ee
The independent number of solutions to (\ref{condf}) tells us how many supersymmetries
the configuration keeps\cite{Justin:susy}.

Again, if the supermatrix $a_{\phi}$ can be diagonalized, we can assume
 \begin{align}
    a_{\phi}\sim \left(\begin{array}{cc}
a_{N\times N}&0\\0&a_{(N-1)\times (N-1)}
\end{array}\right)
\end{align}
with the matrix $a_{N\times N}$ and $a_{(N-1)\times (N-1)}$ to be
\bea
a_{N\times N}\sim diag(\th_1,\cdots,\th_N),\\
a_{(N-1)\times (N-1)}\sim diag(\varphi_{\bar{1}},\cdots,\varphi_{\bar{N}})
\eea
And we have $\th_1>\th_2>\cdots>\th_N,\ \varphi_{\bar{1}}>\varphi_{\bar{2}}>\cdots>\varphi_{\bar{N}}$. The supertraceless condition require
\be
\sum_{i=1}^{N}\th_i-\sum_{\bar{j}=1}^{N-1}\varphi_{\bar{j}}=0
\ee
The supermatrix has $2N-1$ indices: the first N indices will be denoted as $i,j$ and the last $N-1$ indices will be denoted as $\bar{i},\bar{j}$. Then a basis for the fermionic generators can be chosen to be $E_{i\bar{j}}$ and $E_{\bar{i}j}$. Due to the identity\footnote{There is a similar identity for $E_{\bar{i}j}$.}
\be
\exp\big(-2\pi(\sum_{i=1}^N\th_i E_{ii}+\sum_{\bar{j}=1}^{N-1}\varphi_{\bar{j}}E_{\bar{j}\bar{j}})\big)E_{i\bar{j}}\exp\big(2\pi(\sum_{i=1}^N\th_i E_{ii}+\sum_{\bar{j}=1}^{N-1}\varphi_{\bar{j}}E_{\bar{j}\bar{j}})\big)=e^{-2\pi(\th_i-\varphi_{\bar{j}})}E_{i\bar{j}},
\ee
the condition (\ref{condf}) has a solution when there exists $i,\bar{j}$ such that
\be
\th_i-\varphi_{\bar{j}}=i n,\ n\in\mathbb{Z}\label{cri1}
\ee
for periodic boundary condition,
and
\be
\th_i-\varphi_{\bar{j}}=i (n+\frac{1}{2}),\ n\in\mathbb{Z}\label{cri2}
\ee
for anti-periodic boundary condition.

In the case that $a_{\phi}$ cannot be diagonalized, then we need to solve (\ref{condf}) from scratch. In the following section we will
%Or we can solve killing spinor equation from scratch and
find that some black holes do preserve part of supersymmetries discussed here.

\section{Black Holes}

The higher spin black holes are the classical solutions of the flatness equation of motion. To have a smooth geometry,  the holonomy of the gauge potential of the black hole around the thermal circle is required to be in the center of the corresponding group. The holonomy condition has been checked for the black holes in spin 3 \cite{Per kraus:2011}, spin 4 \cite{Tan}, spin $\tilde{4}$, and $G_2$ gravities\cite{truncated}, and it has been extended to the case with supersymmetry\cite{Tan:2012}. In all the cases, it leads to consistent thermodynamics, together with the integrability condition.

 It is interesting to compare the higher spin black holes with the conical defects discussed in the previous sections. For the smooth conical defects, the most important feature is that the holonomy around the spatial circle is trivial,
 \be
 H_{\phi}(A)=\exp{\oint_{spatial\ circle} A}\in\ center.\label{sp}
 \ee
 On the contrary, the higher spin black holes require the holonomy around the thermal circle is trivial,
 \be
 H_{\tau}(A)=\exp{\oint_{thermal\ circle} A}\in\ center.\label{th}
 \ee
  Note that we have shown that when the holonomy around the spatial circle is trivial, the solution is maximally symmetric, while if the spatial holonomy is not trivial, the solution is only partially symmetric. In particular, the solution may keep part of supersymmetries. For the higher spin black hole, its spatial holonomy is nontrivial
  so it is only to be partially supersymmetric.

  In this section, we would like to discuss the black hole solutions in the higher spin supergravity and check if there are supersymmetric ones. First of all, we need to construct suitable higher spin black holes. The black holes we find have not been discussed before hence we will clarify these solutions in detail, including the explicit solutions, and most importantly, how the holonomies around the thermal circle lead to consistent thermodynamics. Next we discuss under what condition, the black hole become supersymmetric. We will use two methods to investigate the issue. The first one is to solve the generalized Killing equation by brute force. The second one is to discuss the relation (\ref{condf}) on the holonomy. It turns out that two methods always lead to the same answer. To simplify the discussion, we work on the higher spin black holes whose explicit entropy forms are feasible. These include the ones in $osp(3|2)$ supergravity and $sl(3|2)$ supergravity.

\subsection{Black Holes in $osp(3|2)$ Supergravity}

The study of the black hole solution in a higher spin supergravity is not much difficult than the bosonic one. To find the black holes in the higher spin supergravity, we only need the bosonic algebra of the theory. However, from the structure of the higher spin algebra, the bosonic part is just the direct sum of decoupled semi-simple Lie algebras. This property is quite like D$_2$ gravity \cite{D2}. One lesson from the $D_2$ gravity is that the total entropy of the black hole can be the sum of the entropies of two decoupled system. The same is true for the black hole in the higher spin supergravity. %The Hence we say that the thermodynamic study of higher spin black holes in higher spin supergravity is no more complex than bosonic case conceptually. Besides the decoupled property of the black hole solution, higher spin supergravity is different from the D$_2$ gravity.
In this subsection, we explore the black holes in $osp(3|2)$ supergravity. This supergravity can be taken the simplest higher spin supergravity as it contains a spin $5/2$ field. %We first find the black hole solutions and then study various properties of the black holes, including its thermodynamics and supersymmetry. We mention that analysis of the solution and the thermodynamics of higher spin black holes in higher spin supergravity is no more complex than the bosonic case. We can illustrate this point briefly.

\subsubsection{Solution}

Let us first consider the solution in the $osp(3|2)$ supergravity. As in the $sl(N)$ gravity, we can choose the highest-weight gauge to set the gauge connection to be
\bea
A=e^{-\rho L_0}a e^{\rho L_0}+L_0 d\rho,\nn\\
\bar{A}=e^{\rho L_0}\bar{a}e^{-\rho L_0}-L_0d\rho\label{connection}
\eea
where
\bea
a&=&(L_1-\mathcal{L}L_{-1}+\mathcal{L}'A_{-1})dx^++(\nu L_{-1}+\sum_{i=-1}^{1}q_i A_i)dx^-,\nn\\
\bar{a}&=&-(L_{-1}-\bar{\mathcal{L}}L_1+\bar{\mathcal{L}}'A_1)dx^--(\bar{\nu}L_1+\sum_{i=-1}^{1}\bar{q}_iA_i)dx^+
\eea
From the equation of motion, we find the solution
\bea
a&=&(L_1-\mathcal{L}L_{-1}+\mathcal{L}'A_{-1})dx^++\mu(\mathcal{L}'L_{-1}+A_1-\mathcal{L}A_{-1})dx^-,\nn\\
\bar{a}&=&-(L_{-1}-\bar{\mathcal{L}}L_1+\bar{\mathcal{L}}'A_1)dx^--\bar{\mu}(A_{-1}-\bar{\mathcal{L}}A_1+\bar{\mathcal{L}}'L_1)dx^+.
\eea
Here we have relabel the parameters $q$ and $\bar{q}$ to be $\mu$ and $\bar{\mu}$, which could be interpreted  as the potentials conjugate to the spin 2 charges $\mathcal{L}'$ and $\bar{\mathcal{L}}'$. Note that this solution is exactly the same as the one in the $D_2$ gravity\cite{D2}. However, the $osp(3|2)$ gravity is different from the $D_2$ gravity in many aspects. First of all, the $osp(3|2)$ gravity has many fermionic degrees of freedom, which connect the decoupled bosonic degrees of freedom. Secondly, the supertrace of the supermatrix M is different from the trace in the $D_2$ gravity. Therefore this $osp(3|2)$ higher spin supergravity may not have a second order formulation as the $D_2$ gravity.

\subsubsection{Thermodynamics}

The thermodynamics of the black holes found in previous section can be studied by obtaining its exact entropy. Now the holomorphic part of the partition function is
 \be
Z=Tr \exp{2\pi k i(\tau\mathcal{L}+\alpha\mathcal{L}')}
\ee
We denote the entropy of the black hole to be $S$, which is a function of $\mathcal{L}$ and $\mathcal{L}'$.  The parameters $\tau,\alpha$ can be related to $\mathcal{L},\mathcal{L}'$ by
\be
\tau=\frac{i}{2\pi k}\frac{\partial S}{\partial \mathcal{L}},\ \alpha=\frac{i}{2\pi k}\frac{\partial S}{\partial \mathcal{L}'}\label{mu}
\ee
Equivalently we may define four other variables $\mathcal{H},\mathcal{K},\gamma,\delta$ by
\bea
\mathcal{H}&\equiv&\mathcal{L}-\mathcal{L}',\ \mathcal{K}\equiv\mathcal{L}+\mathcal{L}',\\
\gamma&\equiv&\tau-\alpha,\ \delta\equiv\tau+\alpha
\eea
which have
\be
\gamma=\frac{i}{\pi k}\frac{\partial S}{\partial\mathcal{H}},\ \delta=\frac{i}{\pi k}\frac{\partial S}{\partial\mathcal{K}}.\label{gamma}
\ee
The holonomy around the thermal circle is $H=e^{\oint}A\equiv e^{\omega}$, with
\begin{align}
\omega=2\pi(a_+\tau-a_-\bar{\tau})=\left(\begin{array}{cc}
\omega_{3\times3}&0\\ 0&\omega_{2\times2}\end{array}
\right).
\end{align}
In the bosonic higher spin gravity, the holonomy has  to be in the center of the corresponding algebra\cite{Per kraus:2011}. In the higher spin supergravity we are considering,  the holonomy is
\begin{align}
H=\left(\begin{array}{cc}
1_{3\times3}&0\\ 0&-1_{2\times2}\end{array}
\right).
\end{align}
In other words, we require the eigenvalues of $\omega_{3\times3}$ to be $2\pi i,0,-2\pi i$ and the ones of $\omega_{2\times2}$ to be $\pi i,-\pi i$. After a short computation, we find
\begin{align}
\omega_{3\times3}=2\pi\left(\begin{array}{ccc}
0&\sqrt{2}\mathcal{H}\gamma\\ \sqrt{2}\gamma&0&\sqrt{2}\mathcal{H}\gamma\\0&\sqrt{2}\gamma&0\end{array}
\right),\ \ \omega_{2\times2}=2\pi\left(\begin{array}{cc}
0&\mathcal{K}\delta\\ \delta&0\end{array}\right).
\end{align}
Hence the holonomy equation becomes
\be
tr\omega_{3\times3}^2=-8\pi^2,\ tr\omega_{2\times2}^2=-2\pi^2.
\ee
From the equation (\ref{gamma}) we find the entropy of the black hole to be
\be
S=\pm\pi k\sqrt{\mathcal{H}}\pm\pi k\sqrt{\mathcal{K}}
\ee
There are four branches of the solutions
 \begin{itemize}
\item{Branch 1: $S=\pi k(\sqrt{\mathcal{H}}+\sqrt{\mathcal{K}})$}
\item{Branch 2: $S=\pi k(\sqrt{\mathcal{H}}-\sqrt{\mathcal{K}})$}
\item{Branch 3: $S=\pi k(-\sqrt{\mathcal{H}}+\sqrt{\mathcal{K}})$}
\item{Branch 4: $S=\pi k(-\sqrt{\mathcal{H}}-\sqrt{\mathcal{K}})$}
\end{itemize}
Note that for each branch, we can find the corresponding $\tau,\alpha$ and then $\mu$. One can explore the phase structure of the black holes as in
\cite{Justin:thermo,Phase}, but we will not include it here. In the following, we will explore the supersymmetries of the solution.
%From
%\bea
%&&\tau=\frac{i}{2\pi k}(\frac{\partial S}{\partial\mathcal{H}}+\frac{\partial S}{\partial\mathcal{K}})=\frac{i}{4}(\pm\frac{1}{\sqrt{\mathcal{H}}}\pm\frac{1}{\sqrt{\mathcal{K}}})\\
%&&\alpha=\frac{i}{2\pi k}(\frac{\partial S}{\partial\mathcal{H}}-\frac{\partial S}{\partial\mathcal{K}})=\frac{i}{4}(\pm\frac{1}{\sqrt{\mathcal{H}}}\mp\frac{1}{\sqrt{\mathcal{K}}})
%\eea
%one got
%\bea
%T&=&\frac{2\pi i}{\tau}=8\pi\frac{\sqrt{\mathcal{H}\mathcal{K}}}{\pm\sqrt{\mathcal{K}}\pm\sqrt{\mathcal{H}}}\label{ospT}\\
%\mu&=&-\frac{\alpha}{\tau}=-\frac{\pm\sqrt{\mathcal{K}}\mp\sqrt{\mathcal{H}}}{\pm\sqrt{\mathcal{K}}\pm\sqrt{\mathcal{H}}}.
%\eea
%In this case it's different from the spin-3 black hole and spin-$\tilde{4}$ black hole in \cite{Justin:thermo}\cite{Phase}, since the chemical potential $\mu$ is dimensionless. Hence if one fix $\mu$, the proportion between $\mathcal{L}$ and $\mathcal{Y}$ is fixed, and we won't do that. We find for branch 1 and branch 4, $\sqrt{\mathcal{H}}=\frac{\mu+1}{1-\mu}\sqrt{\mathcal{K}}$, hence $1\geq\mu\geq -1$. For branch 2 and branch 3, $\sqrt{\mathcal{H}}=\frac{\mu+1}{\mu-1}\sqrt{\mathcal{K}}$, hence $\mu\geq 1$ or $\mu\leq -1$. For all the four branches, there's a relation:
%\be
%\frac{S}{T}=\frac{k}{2(1-\mu^2)}.
%\ee
%One can simply recognize that the only branch with positive $S$ and $T$ is branch 1. And from the expression of $T$(\ref{ospT}), the temperature does not have a upper limit. Hence there's no phase transition, the system will always stay on branch 1.

\subsubsection{Supersymmetry I}

In a supergravity, it is important to know  how many supersymmetries the solution preserves.  The supersymmetric condition is a generalized Killing spinor equation
\be
d\epsilon+[A,\epsilon]=0\label{spinor}
\ee
where the spinor $\epsilon$ can be expanded as
\be
\epsilon=\eta_r R_r+\k_s Z_s
\ee
with $r=-\frac{3}{2},-\frac{1}{2},\frac{1}{2},\frac{3}{2}$ and $s=-\frac{1}{2},\frac{1}{2}$ for $osp(3|2)$ case. We notice that in the paper \cite{Tan:2012}, only  the spin $1/2$ generators were included in the expansion. Nevertheless, we feel that it is more reasonable to include the higher spin fermonic generators, as did in \cite{Justin:susy,Hikida:3}.

From the form of the connection (\ref{connection}) and the $\rho$ component of the Killing spinor equation (\ref{spinor}), $\epsilon$ can be cast into the form
\be
\epsilon=e^{-\rho L_0}\epsilon_0 e^{\rho L_0}
\ee
where $\epsilon_0$ is a function independent of $\rho$. The $+$ component of the  Killing spinor equation can be expanded as
\bea
&&\partial_+\eta_{\frac{3}{2}}+\eta_{\frac{1}{2}}=0,\nn\\
&&\partial_+\eta_{\frac{1}{2}}+2\eta_{-\frac{1}{2}}+3\mathcal{L}\eta_{\frac{3}{2}}-\mathcal{L}'\eta_{\frac{3}{2}}=0,\nn\\
&&\partial_+\eta_{-\frac{1}{2}}+3\eta_{-\frac{3}{2}}+2\mathcal{L}\eta_{\frac{1}{2}}-\frac{2}{3}\mathcal{L}'\eta_{\frac{1}{2}}+\frac{4}{3}\mathcal{L}'\k_{\frac{1}{2}}=0,\label{+}\\
&&\partial_+\eta_{-\frac{3}{2}}+\mathcal{L}\eta_{-\frac{1}{2}}-\frac{1}{3}\mathcal{L}'\eta_{-\frac{1}{2}}+\frac{4}{3}\mathcal{L}'\k_{-\frac{1}{2}}=0,\nn\\
&&\partial_+\k_{\frac{1}{2}}+\k_{-\frac{1}{2}}-2\mathcal{L}'\eta_{\frac{3}{2}}=0,\nn\\
&&\partial_+\k_{-\frac{1}{2}}+\mathcal{L}\k_{\frac{1}{2}}-\frac{2}{3}\mathcal{L}'\eta_{\frac{1}{2}}-\frac{5}{3}\mathcal{L}'\k_{\frac{1}{2}}=0.\nn
\eea
The $-$ component of the Killing spinor equation can be expanded as
\bea
&&\partial_-\eta_{\frac{3}{2}}+\frac{1}{3}\mu\eta_{\frac{1}{2}}+\frac{4}{3}\mu\k_{\frac{1}{2}}=0,\nn\\
&&\partial_-\eta_{\frac{1}{2}}-3\mu\mathcal{L}'\eta_{\frac{3}{2}}+\frac{2}{3}\mu\eta_{-\frac{1}{2}}+\mu\mathcal{L}\eta_{\frac{3}{2}}+\frac{4}{3}\mu\k_{-\frac{1}{2}}=0,\nn\\
&&\partial_-\eta_{-\frac{1}{2}}-2\mu\mathcal{L}'\eta_{\frac{1}{2}}+\mu\eta_{-\frac{3}{2}}+\frac{2}{3}\mu\mathcal{L}\eta_{\frac{1}{2}}-\frac{4}{3}\mu\mathcal{L}\k_{\frac{1}{2}}=0,\nn\\
&&\partial_-\eta_{-\frac{3}{2}}-\mu\mathcal{L}'\eta_{-\frac{1}{2}}+\frac{1}{3}\mu\mathcal{L}\eta_{-\frac{1}{2}}-\frac{4}{3}\mu\mathcal{L}\k_{\frac{1}{2}}=0,\label{-}\\
&&\partial_-\k_{\frac{1}{2}}-\frac{2}{3}\mu\eta_{-\frac{1}{2}}+\frac{5}{3}\mu\k_{-\frac{1}{2}}+2\mu\mathcal{L}\eta_{\frac{3}{2}}=0,\nn\\
&&\partial_-\k_{-\frac{1}{3}}-\mu\mathcal{L}'\k_{\frac{1}{2}}-2\mu\eta_{-\frac{3}{2}}+\frac{2}{3}\mu\mathcal{L}\eta_{\frac{3}{2}}+\frac{5}{3}\mu\mathcal{L}\k_{\frac{1}{2}}=0.\nn
\eea

Let us first set $\mathcal{L}'$ to be zero. In this case we have $\mu=0$\footnote{We are interested in Branch 1, then the zero $\mathcal{L}$' leads to zero $\mu$. A general feature of higher spin black hole is that in some branches\cite{Phase} other than Branch 1 even if the higher spin charge $\mathcal{W}$ vanishes  the corresponding $\mu$ can be non-zero. We will not include these cases here.}.
%Strictly speaking, the solution is just a black hole anymore. Nevertheless we take it as the toy model to test our methods.
 From the  equations (\ref{-}) we find that $\eta_r,\k_s$ are independent of $x^-$. Then the  equations (\ref{+}) become
\bea
&&(\partial_+^2-\mathcal{L})\k_{\frac{1}{2}}=0,\nn\\
&&(\partial_+^2-\mathcal{L})(\partial_+^2-9\mathcal{L})\eta_{\frac{3}{2}}=0.\label{killing}
\eea
The other components can be obtained easily. If $\mathcal{L}\not=0$, the general solution should be
\bea
\k_{\frac{1}{2}}&=&c_1 e^{\sqrt{\mathcal{L}}x^+}+c_2 e^{-\sqrt{\mathcal{L}}x^+}\nn\\
\eta_{\frac{3}{2}}&=&d_1 e^{3\sqrt{\mathcal{L}}x^+}+d_2 e^{\sqrt{\mathcal{L}}x^+}+d_3 e^{-\sqrt{\mathcal{L}} x^+}+d_4 e^{-3\sqrt{\mathcal{L}}x^+}.
\eea
Then for general $\mathcal{L}>0$, the Killing spinor is neither periodic nor anti-periodic, so a general $BTZ$ black hole breaks the supersymmetries completely. However, for $\mathcal{L}<0$, the Killing spinor can satisfy the periodic or anti-periodic condition. If we restrict ourselves in the region $-\frac{1}{4}\le\mathcal{L}<0$, then when $\mathcal{L}=-\frac{1}{4}$ the spnior satisfy anti-periodic boundary condition. Since $\mathcal{L}=-\frac{1}{4},\mathcal{L}'=0$ corresponds to global $AdS_3$, we conclude that the global $AdS_3$ preserve all the supersymmetries. For the anti-holomorphic sector, we have the same conclusion. We use (6,6) to denote the total supersymmetries since there are six Killing spinors in each sector\footnote{Here we use $(p,\bar{p})$ to denote the number of supersymmetries. $p$($\bar{p}$) is the number of the independent killing spinors of the holomorphic (anti-holomorphic) part.}  .

However, in the range $-\frac{1}{4}\le\mathcal{L}<0$, there are two other interesting cases. If $\mathcal{L}=-\frac{1}{9}$, we find that we can set
\be
c_1=c_2=d_2=d_3=0,
\ee
and $d_1, d_4$ to be arbitrary constant.
Then the configuration can preserve two supersymmetries, and the corresponding Killing spinors satisfy periodic boundary condition.
 If $\mathcal{L}=-\frac{1}{36}$, we may set
 \be
 c_1=c_2=d_2=d_3=0
 \ee
 and $d_1,d_4$ to be arbitrary constant, such that the configuration can preserve two supersymmetries with the corresponding Killing spinors satisfying anti-periodic boundary condition.  These supercharges are coming from the spin-$5/2$ components hence are different from the conventional supercharges. %It relates to our generalization of supercharge to the higher spin case.

Another interesting case is when $\mathcal{L}=\mathcal{L}'=0$. Then the general solution of Eq. (\ref{killing}) is
\bea
\k_{\frac{1}{2}}&=&c_1 x^++c_2,\nn\\
\eta_{\frac{3}{2}}&=&d_1 (x^+)^3+d_2 (x^+)^2+d_3(x^+)+d_4.
\eea
Only when $c_1=d_1=d_2=d_3=0$, the Killing spinor preserve periodic boundary condition.
This configuration was called extreme BTZ black hole in the literature.

The results can be summarized as follows
\begin{enumerate}
\item
For global $AdS_3$, $\mathcal{L}=\bar{\mathcal{L}}=-\frac{1}{4}$, it preserves (6,6) supersymmetry.
\item
For $\mathcal{L}=-\frac{1}{4},\bar{\mathcal{L}}=-\frac{1}{9}$, it preserves (6,2) supersymmetry.
\item
For $\mathcal{L}=-\frac{1}{9},\bar{\mathcal{L}}=-\frac{1}{4}$, it preserves (2,6) supersymmetry.
\item
For $\mathcal{L}=\bar{\mathcal{L}}=-\frac{1}{9}$, it preserves (2,2) supersymmetry.
\item
For $\mathcal{L}=-\frac{1}{4}, \bar{\mathcal{L}}=-\frac{1}{36}$, it preserves (6,2) supersymmetry.
\item
For $\mathcal{L}=-\frac{1}{36}, \bar{\mathcal{L}}=-\frac{1}{4}$, it preserves (2,6) supersymmetry.
\item
For $\mathcal{L}=\bar{\mathcal{L}}=-\frac{1}{36}$, it preserves (2,2) supersymmetry.
\item
For $\mathcal{L}=-\frac{1}{9},\bar{\mathcal{L}}=-\frac{1}{36}$ or $\mathcal{L}=-\frac{1}{36},\bar{\mathcal{L}}=-\frac{1}{9}$, it preserves (2,2) supersymmetry.
\item
For massless BTZ black hole, $\mathcal{L}=\mathcal{L}'=\bar{\mathcal{L}}=\bar{\mathcal{L}}'=0$, it preserves (2,2) supersymmetry.
\item
For extreme BTZ black hole with nonzero mass, $\mathcal{L}=\mathcal{L}'=0, \bar{\mathcal{L}}\not=0,\bar{\mathcal{L}}'=0$ or $\mathcal{L}\not=0,\mathcal{L}'=0,\bar{\mathcal{L}}=\bar{\mathcal{L}}'=0$, it preserves (2,0) or (0,2) supersymmetry.
\end{enumerate}

The solutions 2 to 8 listed above are not exactly the black holes. And actually they are not smooth conical defects we discussed before. They do preserve some supersymmetries, but are not smooth. %({\bf symmetries?}) %that we know litte about the solution from 2 to 8 besides they preserve some fermionic symmetries. The are neither the smooth
%conical defects, nor higher spin black holes. Maybe they are not allowed from the creterion of smoothness (\ref{sp}), (\ref{th}).

Next we turn to the black hole with a nonvanishing spin 2 charge $\mathcal{L}'\not=0$.
As the general solution to the Killing spinor equation would be very complicated, here we are satisfied with searching for the constant solutions. These solutions satisfy the periodic boundary condition in the $\phi$ direction so the problem reduce to search for non-zero constant solution of the equations (\ref{+}) and (\ref{-}).
%\end{enumerate}

Let us first consider the equations (\ref{+}). These equations  reduce to a set of linear equations when the solutions are constants. They have non-zero solutions if and only if the determinant of the matrix $M_+$
\begin{align}
M_+=\left(\begin{array}{cccccc}
0&1&0&0&0&0\\3\mathcal{L}-\mathcal{L}'&0&2&0&0&0\\0&\frac{1}{3}(6\mathcal{L}-2\mathcal{L}')&0&3&\frac{4\mathcal{L}'}{3}&0\\
0&0&\frac{1}{3}(3\mathcal{L}-\mathcal{L}')&0&0&\frac{4\mathcal{L}'}{3}\\-2\mathcal{L}'&0&0&0&0&1\\0&-\frac{2\mathcal{L}'}{3}&0&0&\frac{1}{3}(3\mathcal{L}-5\mathcal{L}')&0\end{array}
\right)
\end{align}
is zero:
\be
det(M_+)=-(3\mathcal{L}-5\mathcal{L}')^2(\mathcal{L}+\mathcal{L}')=0.\label{+det}
\ee
From the  Killing spinor equations (\ref{-}), the characteristic matrix $M_-$ is
\begin{align}
M_-=\mu \left(\begin{array}{cccccc}
0&\frac{1}{3}&0&0&\frac{4}{3}&0\\\mathcal{L}-3\mathcal{L}'&0&\frac{2}{3}&0&0&\frac{4}{3}\\0&\frac{1}{3}(2\mathcal{L}-6\mathcal{L}')&0&1&-\frac{4\mathcal{L}}{3}&0\\
0&0&\frac{1}{3}(\mathcal{L}-3\mathcal{L}')&0&0&-\frac{4\mathcal{L}}{3}\\2\mathcal{L}&0&-\frac{2}{3}&0&0&\frac{5}{3}\\0&\frac{2\mathcal{L}}{3}&0&-2&\frac{1}{3}(5\mathcal{L}-3\mathcal{L}')&0\end{array}
\right)
\end{align}
and its determinant is
\be
det(M_-)=-\mu^6(3\mathcal{L}-5\mathcal{L}')^2(\mathcal{L}+\mathcal{L}').\label{-det}
\ee
For the equation (\ref{+det}) to be zero, there are two cases:
\begin{enumerate}
\item $\mathcal{L}=\frac{5\mathcal{L}'}{3}$. Then the equations (\ref{+}) have the non-zero solutions
    \be
    \eta_1=0,\ \eta_{-1}=-\k_{-1},\ \k_{-1}=2\eta_3\mathcal{L}',\ \eta_{-3}=-\frac{4\k_1\mathcal{L}'}{9},\ \eta_3=C_1\ \k_1=C_2\label{solu1}
    \ee
    where $C_1, C_2$ are constants. Though (\ref{-det}) is zero for this solution, the solution (\ref{solu1}) does not satisfy the equations (\ref{-}). Thus in general there is no supersymmetry for this configuration. However, there is one exception. That is to set $\mu$ to be zero. Then the equations (\ref{-}) are satisfied. This can be achieved by $\bar{\tau}=\infty$, or equivalently, $\bar{\mathcal{L}}=\bar{\mathcal{L}}'$ or  $\bar{\mathcal{L}}=-\bar{\mathcal{L}}'$. In the case that $\bar{\mathcal{L}}=\bar{\mathcal{L}}'$ or  $\bar{\mathcal{L}}=-\bar{\mathcal{L}}'$, the anti-holomorphic part of the black hole is extreme. So the holomorphic part of the black hole solution will preserve 2 supersymmetries for these extreme black holes.
\item $\mathcal{L}=-\mathcal{L}'$. Then the equations (\ref{+}) have the non-zero solution
    \be
    \eta_1=\eta_{-3}=\k_{1}=0,\ \eta_{-1}=\k_{-1}=-2\eta_3\mathcal{L},\ \eta_3=C,
    \ee
    where $C$ is a constant. For this configuration, the equations in (\ref{-}) are indeed satisfied. In general, such kinds of configuration preserve 1 supersymmetry in the holomorphic part.
\end{enumerate}
For the antiholomorphic sector,  the analysis is similar. Combining the results in two sectors, we list the the configurations that preserve some supersymmetries for non-vanishing spin 2 charge $\mathcal{L}'$ in the following table

\begin{center}
\begin{tabular}{|c|c|}
\hline
Configuration&supersymmetries \\\hline
$\mathcal{L}=\frac{5}{3}\mathcal{L}',\ \bar{\mathcal{L}}=\bar{\mathcal{L}}'$&(2,0)\\\hline
$\mathcal{L}=\mathcal{L}',\ \bar{\mathcal{L}}=\frac{5}{3}\bar{\mathcal{L}}'$&(0,2)\\\hline
$\mathcal{L}=\frac{5}{3}\mathcal{L}',\ \bar{\mathcal{L}}=-\bar{\mathcal{L}}'$&(2,1)\\\hline
$\mathcal{L}=-\mathcal{L}',\ \bar{\mathcal{L}}=\frac{5}{3}\bar{\mathcal{L}}'$&(1,2)\\\hline
$\mathcal{L}=-\mathcal{L}',\ -\bar{\mathcal{L}}<\bar{\mathcal{L}}'\le\bar{\mathcal{L}}, \bar{\mathcal{L}}'\not=\frac{3}{5}\bar{\mathcal{L}}$&(1,0)\\\hline
$-\mathcal{L}<\mathcal{L}'\le\mathcal{L},\mathcal{L}'\not=\frac{3}{5}\mathcal{L}, \bar{\mathcal{L}}=-\bar{\mathcal{L}}'$&(0,1)\\\hline
$\mathcal{L}=-\mathcal{L}',\ \bar{\mathcal{L}}=-\bar{\mathcal{L}}'$&(1,1)\\\hline
\end{tabular}
\end{center}
 We notice that all the solutions are extreme in at least one sector. Moreover, there is a special point $\mathcal{L}=\frac{5}{3}\mathcal{L}'$ where the supersymmetry get enhanced. This configuration  is mysterious for us. It is better to have a deeper understanding. Another remarkable point  is that we only collect the configurations that have some constant Killing spinors. To search for the configurations that have non-constant Killing spinors, we need to solve the Killing spinor equations, which would be quite involved. However, in the following, we can answer this question from the honolomy equation (\ref{condf}).

\subsubsection{Supersymmetry II}

Now we search for the supersymmetric black holes from the spatial holonomy condition (\ref{condf}), without solving the Killing spinor equations.
As the discussion before, we first set $\mathcal{L}'$ and $\bar{\mathcal{L}}'$ to be zero.  Using the symbol $\th_i$, $\varphi_{\bar{j}}$ to denote the eigenvalues of $so(3)$ and $sp(2)$, we find
\bea
\th_1&=&2\sqrt{\mathcal{L}},\ \th_2=0,\ \th_3=-2\sqrt{\mathcal{L}}, \\
\varphi_{\bar{1}}&=&\sqrt{\mathcal{L}},\ \varphi_{\bar{2}}=-\sqrt{\mathcal{L}}.
\eea

When $\mathcal{L}>0$, it is a general non-extreme $BTZ$ black hole. In this case, as the conditions (\ref{cri1}) and (\ref{cri2}) cannot be satisfied, the configuration breaks all the supersymmetry.

When $\mathcal{L}<0$,  the supersymmetry enhancement condition is given by (\ref{cri1}) or (\ref{cri2}). We first consider periodic boundary condition (\ref{cri1}). This can be achieved in two cases.
\begin{enumerate}
\item $\th_1-\varphi_{\bar{1}}=i n, n\in\mathbb{Z}$, which  gives us that
\be
\mathcal{L}=-n^2,\ n\in\mathbb{Z}^+.\label{susyenhance1}
\ee
The negative integer $n$ give the same $\mathcal{L}$ as a positive $n$ and the case $n=0$ is excluded by the condition $\mathcal{L}<0$. %All the solutions with (\ref{susyenhance1}) will preserve 2 fermionic symmetry as there are two
If we restrict ourselves in the range $\mathcal{L}\ge-\frac{1}{4}$, then there is no solution for this case.
\item $\th_1-\varphi_{\bar{2}}=i n,n\in\mathbb{Z}$,  which leads to
\be
\mathcal{L}=-\frac{1}{9}n^2,\ n\in\mathbb{Z}^+.\label{susyenhance2}
\ee
In the range $-\frac{1}{4}\le\mathcal{L}<0$, there is only one solution
\be
\mathcal{L}=-\frac{1}{9},\ n=1.
\ee
This configuration preserve periodic boundary condition. Since there are two pairs of $\th_i-\varphi_{\bar{j}}$($i=1,\bar{j}=2$ and $i=3,\bar{j}=1$) satisfying the condition (\ref{cri1}), it preserve two fermionic symmetries. All these results are the same as the ones we found before.
\end{enumerate}

We can also consider the anti-periodic boundary condition (\ref{cri2}), which can be satisfied in two cases:
\begin{enumerate}
\item $\th_1-\varphi_{\bar{1}}=i (n+\frac{1}{2}), n\in\mathbb{Z}$. It leads to
\be
\mathcal{L}=-(n+\frac{1}{2})^2,\ n\in\mathbb{N}.\label{susyenhance3}
\ee
The unique solution is
\be
\mathcal{L}=-\frac{1}{4},\ n=0
\ee
in the range $-\frac{1}{4}\le\mathcal{L}<0$, leading to the global AdS$_3$. Since for arbitrary $i,\bar{j}$, the condition (\ref{cri2}) can be satisfied,  the configuration actually has maximal supersymmetries. The result is the same  as we found before.
\item $\th_1-\varphi_{\bar{3}}=i (n+\frac{1}{2}), n\in\mathbb{Z}$. It leads to
\be
\mathcal{L}=-\frac{1}{9}(n+\frac{1}{2})^2,\ n\in\mathbb{N}.\label{susyenhance3}
\ee
There are two configurations in the range $-\frac{1}{4}\le\mathcal{L}<0$,
\begin{enumerate}
\item $\mathcal{L}=-\frac{1}{36},\ n=0$, which
preserves two fermionic symmetries and anti-periodic boundary condition.
\item $\mathcal{L}=-\frac{1}{4},\ n=1$, which
is the global AdS$_3$.
\end{enumerate}
\end{enumerate}

Now we want to set $\mathcal{L}=0$. This is a special configuration as it corresponds to extreme $BTZ$ black hole. The eigenvalues of $a_+$ are
\be
\th_i=0,\ \varphi_{\bar{j}}=0, \ \forall i\in \{1,2,3\}, \ \forall \bar{j}\in\{\bar{1},\bar{2}\}
\ee
 In this case $a_+$ cannot be diagonalized. We need to solve the Killing spinor equation as we have done in the previous subsubsection.

 Let us turn to the case $\mathcal{L}'\not=0,\ \bar{\mathcal{L}}'\not=0$. From the discussion of thermodynamics, we have $\mathcal{L}>0$ and  $-\mathcal{L}\le\mathcal{L}'\le\mathcal{L}$, otherwise the solution is not a higher spin black hole. In this case we can find the following eigenvalues of $a_{\phi}$,
\bea
\th_1&=&2(1-\mu)\sqrt{\mathcal{L}-\mathcal{L}'},\ \th_2=0,\ \th_3=-2(1-\mu)\sqrt{\mathcal{L}-\mathcal{L}'},\\
\varphi_{\bar{1}}&=&(1+\mu)\sqrt{\mathcal{L}+\mathcal{L}'},\ \varphi_{\bar{2}}=-(1+\mu)\sqrt{\mathcal{L}+\mathcal{L}'}.
\eea
Note that the extreme case is special, as when $\mathcal{L}'=\pm\mathcal{L}$, the eigenvalue is degenerate, we need to solve the Killing spinor equation from scratch. And this is the same as we discussed previously. Therefore we focus our attention to the case $\mathcal{L}'\not=\pm\mathcal{L}$. Since all the eigenvalues are real, the anti-periodic boundary condition (\ref{cri2}) is impossible and the periodic boundary condition (\ref{cri1}) can only be satisfied by setting $n=0$, for some $i,\ \bar{j}$. Then we find that
\be
2(1-\mu)\sqrt{\mathcal{L}-\mathcal{L}'}=\pm(1+\mu)\sqrt{\mathcal{L}+\mathcal{L}'}.\label{susyenhance5}
\ee
 If we substitute the definition of $\mu=\frac{\alpha}{\bar{\tau}}$ into the equation, then we find that the condition
(\ref{susyenhance5}) can always be satisfied by suitable choice of $\mathcal{L},\mathcal{L}',\bar{\mathcal{L}},\bar{\mathcal{L}}'$. Thus the configuration can preserve two fermionic symmetries. This conclusion sounds different from the one in previous subsubsection. However, there is no contradiction.  In the last subsection we have just solved the Killing spinor equation for the constant solution case.  To have a  constant solution, we require $\mu=0$ and this is achieved by setting $\bar{\mathcal{L}}'=\pm\bar{\mathcal{L}}$. Under this condition, the solution of (\ref{susyenhance5}) leads to
\be
\mathcal{L}=\frac{5}{3}\mathcal{L}',
\ee
 which is in exact agreement with the result got before. However if we do not require a constant solution, then the equation (\ref{susyenhance5}) tells us the whole story.

 Moreover, there is another advantage to work with (\ref{susyenhance5}). It is
an algebraic equation rather than a differential equation, as a Killing spinor equation is. %It is gauge invariant so independent of the coordinates.

 The condition (\ref{susyenhance5}) is interesting because it can be satisfied  without taking extreme limit. This is very different from the conventional supergravity, in which the supersymmetries enhancement always happens for extremal black holes. Now, in the higher spin supergravity, we find that even for non-extreme higher spin black hole, the nonconstant Killing spinors exist.

\subsection{Black Holes in $sl(3|2)$ Supergravity}
The $osp(3|2)$ black hole has no spin $s>2$ hair, so we now consider the black holes with higher spin hair in $sl(3|2)$ supergravity in this section.

\subsubsection{Thermodynamics}

Since the $U(1)$ part is decoupled from the other bosonic generators, we consider the solution with vanishing $U(1)$ charge for simplicity. The solution can be parameterized as
\bea
a_+&=&L_1-\mathcal{L}L_{-1}+\frac{\mathcal{W}}{2}W_{-2}-\mathcal{Y}A_{-1},\nn\\
a_-&=&\nu L_{-1}+q W_2+q_0 W_0+q_{-2}W_{-2}+p A_1+p_{-1}A_{-1},\label{sl32}
\eea
where $\mathcal{L},\mathcal{Y}$ are two spin 2 charges, $\mathcal{W}$ is the spin 3 charge and the constant $\nu,q_0,q_{-2},p_{-1}$ are
\bea
&&\nu=-\mathcal{Y}p-2q\mathcal{W},\ q_0=-2q\mathcal{L}-2q\mathcal{Y},\nn\\
&&p_{-1}=-\mathcal{L}p-2q\mathcal{W},\ q_{-2}=\frac{1}{2}[p\mathcal{W}+2q(\mathcal{L}+\mathcal{Y})^2)].\label{sl32p}
\eea
The holonomy is
\be
\omega=2\pi (a_+\tau-a_-\bar{\tau})
\ee
In terms of
\be
q=-\alpha_3/\bar{\tau}, p=-\alpha_2/\bar{\tau},\label{pq}
\ee
 the holonomy can be written as
\begin{align}
\omega=\left(\begin{array}{cc}
\omega_{3\times3}&0\\ 0&\omega_{2\times2}\end{array}
\right)
\end{align}
with
\begin{align}
\omega_{3\times3}=\left(\begin{array}{ccc}
-\frac{8}{3}\alpha_3\pi(\mathcal{L}+\mathcal{Y})&2\sqrt{2}\pi(4\alpha_3\mathcal{W}+(\alpha_2+\tau)(\mathcal{L}+\mathcal{Y}))&4\pi(\mathcal{W}(\alpha_2+\tau)+2\alpha_3(\mathcal{L}+\mathcal{Y})^2)\\
2\sqrt{2}\pi(\alpha_2+\tau)&\frac{16}{3}\alpha_3\pi(\mathcal{L}+\mathcal{Y})&2\sqrt{2}\pi(4\alpha_3\mathcal{W}+(\alpha_2+\tau)(\mathcal{L}+\mathcal{Y})\\
8\alpha_3\pi&2\sqrt{2}\pi(\alpha_2+\tau)&-\frac{8}{3}\alpha_3\pi(\mathcal{L}+\mathcal{Y})\end{array}
\right),
\end{align}
\begin{align}
\omega_{2\times2}=\left(\begin{array}{cc}
0&2\pi(\tau-\alpha_2)(\mathcal{L}-\mathcal{Y})\\
2\pi(\tau-\alpha_2)&0\end{array}
\right).
\end{align}
 Similar to the $osp(3|2)$ case, we use the holonomy condition that the eigenvalues of $\omega_{3\times3}$ are $2i\pi,0,-2i\pi$ and the eigenvalues of $\omega_{2\times2}$ are $i\pi,-i\pi$. We can also determine $\tau,\alpha_2,\alpha_3$ by
\be
\tau=\frac{i}{2\pi k}\frac{\partial{S}}{\partial{\mathcal{L}}},\ \alpha_2=\frac{i}{2\pi k}\frac{\partial{S}}{\partial{\mathcal{Y}}},\ \alpha_3=\frac{i}{2\pi k}\frac{\partial{S}}{\partial{\mathcal{W}}}.\label{tal}
\ee
Here $S$ is the entropy of the black hole. It is a function of the charges $\mathcal{L},\mathcal{Y}$ and $\mathcal{W}$. It is more convenient to redefine the charges to be $\mathcal{L}+\mathcal{Y}$,$\mathcal{L}-\mathcal{Y}$ and $\mathcal{W}$. It is easy to see that the holonomy $\omega_{3\times3}$ is only dependent of $\mathcal{L}+\mathcal{Y}$ and $\mathcal{W}$, $\alpha_2+\tau$ and $\alpha_3$. And as
\be
\alpha_2+\tau=\frac{i}{\pi k}\frac{\partial{S}}{\partial(\mathcal{L}+\mathcal{Y})},
\ee
 $\omega_{3\times 3}$ depends on $\mathcal{L}+\mathcal{Y}$ and $\mathcal{W}$ and their corresponding potentials. Similary, $\omega_{2\times2}$ depends on $\mathcal{L}-\mathcal{Y}$ and its potential.  Therefore we can cast the entropy function to be
\be
S(\mathcal{L},\mathcal{Y},\mathcal{W})=S_1(\mathcal{L}+\mathcal{Y},\mathcal{W})+S_2(\mathcal{L}-\mathcal{Y}).\label{entropyosp}
\ee
Then we can solve the holonomy equations by
\bea
S_i&=&\pi k \sqrt{\mathcal{L}-\mathcal{Y}}+\pi k \sqrt{\mathcal{L}+\mathcal{Y}}\cos\frac{1}{3}(\arcsin\frac{3\sqrt{3}z}{4}+(i-1) \pi), 1\le i\le 6\\
S_{i+6}&=&-\pi k\sqrt{\mathcal{L}-\mathcal{Y}}+\pi k \sqrt{\mathcal{L}+\mathcal{Y}}\cos\frac{1}{3}(\arcsin\frac{3\sqrt{3}z}{4}+(i-1) \pi), 1\le i\le 6
\eea
where the parameter $z$ is
\be
z=\frac{\mathcal{W}}{(\mathcal{L}+\mathcal{Y})^{\frac{3}{2}}}.
\ee
Note that there are 12 branches of solutions in the holomorphic part. And for a general $sl(3|2)$ higher spin black hole with vanishing $U(1)$ charge there  would be  $(12\times12=)$144 branches. It could be interesting to study  the phase structure of the $sl(3|2)$ black hole as in \cite{Phase}.

The entropy of the black hole in $sl(3|2)$ is actually a sum of the entropies of the spin 2 BTZ black hole and the spin 3 black hole. This is due to the decoupling of two sets of the bosonic generators. As there are 2 branches for the BTZ black hole and 6 branches for the spin 3 black holes\cite{Justin:thermo,Phase}, there are totally 12 branches in each sector.

\subsubsection{Supersymmetry I}

As the $osp(3|2)$ we studied above, we may solve the Killing spinor equation directly to read the conditions for supersymmetry
\be
d\epsilon+[A,\epsilon]=0.
\ee
Here the spinor $\epsilon$ still has the form
\be
\epsilon=e^{-\rho L_0}\epsilon_0e^{\rho L_0}
\ee
where $\epsilon_0$ is independent of the $\rho$ coordinate. It can be expanded as
\bea
\epsilon_0&=&\eta_{1} G_{\frac{1}{2}}+\eta_{2}G_{-\frac{1}{2}}+\eta_{3}H_{\frac{1}{2}}+\eta_{4}H_{-\frac{1}{2}}+\eta_{5}S_{\frac{3}{2}}+\eta_{6}S_{\frac{1}{2}}\nn\\
&&+\eta_7S_{-\frac{1}{2}}+\eta_8S_{-\frac{3}{2}}+\eta_{9}T_{\frac{3}{2}}+\eta_{10}T_{\frac{1}{2}}+\eta_{11}T_{-\frac{1}{2}}+\eta_{12}T_{-\frac{3}{2}}.
\eea
We use the same symbol $\eta$ to simplify notation, but one should keep in mind that they are the coefficients in front of different types of fermonic generators. The $``+''$ component of the Killing spinor equation gives us
\bea
&&\partial_+\eta_{1}+\eta_{2}+2\eta_{5}\mathcal{Y}=0,\nn\\
&&\partial_+\eta_{2}+\frac{1}{3}(3\eta_{1}\mathcal{L}-6\eta_{5}\mathcal{W}+5\eta_{1}\mathcal{Y}+2\eta_{6}\mathcal{Y})=0\nn\\
&&\partial_+\eta_{3}+\eta_{4}-2\eta_{9}\mathcal{Y}=0\nn\\
&&\partial_+\eta_{4}+\frac{1}{3}(3\eta_{3}\mathcal{L}-6\eta_{9}\mathcal{W}-2\eta_{10}\mathcal{Y}+5\eta_3\mathcal{Y})=0,\nn\\
&&\partial_+\eta_{5}+\eta_{6}=0\nn\\
&&\partial_+\eta_{6}+2\eta_{7}+3\eta_{5}\mathcal{L}+\eta_{5}\mathcal{Y}=0\label{sl+}\\
&&\partial_+\eta_{7}+\frac{1}{3}(9\eta_{8}+6\eta_{6}\mathcal{L}-6\eta_{5}\mathcal{W}-4\eta_{1}\mathcal{Y}+2\eta_{6}\mathcal{Y})=0\nn\\
&&\partial_+\eta_{8}+\frac{1}{3}(3\eta_{7}\mathcal{L}+4\eta_{1}\mathcal{W}-2\eta_{6}\mathcal{W}-4\eta_{2}\mathcal{Y}+\eta_{7}\mathcal{Y})=0\nn\\
&&\partial_+\eta_{9}+\eta_{10}=0\nn\\
&&\partial_+\eta_{10}+2\eta_{11}+3\eta_{9}\mathcal{L}+\eta_{9}\mathcal{Y}=0\nn\\
&&\partial_+\eta_{11}+\frac{1}{3}(9\eta_{12}+6\eta_{10}\mathcal{L}+6\eta_{9}\mathcal{W}+2\eta_{10}\mathcal{Y}+4\eta_{3}\mathcal{Y})=0,\nn\\
&&\partial_+\eta_{12}+\frac{1}{3}(3\eta_{11}\mathcal{L}+2\eta_{10}\mathcal{W}+4\eta_{3}\mathcal{W}+\eta_{11}\mathcal{Y}+4\eta_{4}\mathcal{Y})=0\nn
\eea
The $``-''$ component of Killing spinor gives us
\bea
&&\partial_-\eta_{1}+\frac{1}{3}(5\eta_{2}p-2\eta_{7}p+6\eta_{5}\mathcal{L}p+12\eta_{8}q-4\eta_{6}\mathcal{L}q+12\eta_{5}\mathcal{W}q-4\eta_{6}\mathcal{Y}q)=0\nn\\
&&\partial_-\eta_{2}+\frac{1}{3}(-6\eta_{8}p+5\eta_{1}\mathcal{L}p+2\eta_{6}\mathcal{L}p-6\eta_{5}\mathcal{W}p+4\eta_{7}\mathcal{L}q-12\eta_{5}\mathcal{L}^2q+\nn\\
&&\ \ \ \ \ \ \ \ \ \ \ 16\eta_{1}\mathcal{W}q+4\eta_{6}\mathcal{W}q+3\eta_{1}\mathcal{Y}p+4\eta_{7}\mathcal{Y}q-24\eta_{5}\mathcal{L}\mathcal{Y}q-12\eta_{5}\mathcal{Y}^2q)=0\nn\\
&&\partial_-\eta_{3}+\frac{1}{3}(2\eta_{11}p+5\eta_{4}p-6\eta_{9}\mathcal{L}p+12\eta_{12}q-4\eta_{10}\mathcal{L}q-12\eta_{9}\mathcal{W}q-4\eta_{10}\mathcal{Y}q)=0\nn\\
&&\partial_-\eta_{4}+\frac{1}{3}(6\eta_{12}p-2\eta_{10}\mathcal{L}p+5\eta_{3}\mathcal{L}p-6\eta_{9}\mathcal{W}p+4\eta_{11}\mathcal{L}q-12\eta_{9}\mathcal{L}^2q-\nn\\
&&\ \ \ \ \ \ \ \ \ \ \
4\eta_{10}\mathcal{W}q+16\eta_{3}\mathcal{W}q+3\eta_{3}\mathcal{Y}p+4\eta_{11}\mathcal{Y}q-24\eta_{9}\mathcal{L}\mathcal{Y}q-12\eta_{9}\mathcal{Y}^2q)=0\nn\\
&&\partial_-\eta_{5}+\frac{1}{3}(4\eta_{1}p+\eta_{6}p-8\eta_{2}q-4\eta_{7}q+4\eta_{5}\mathcal{L}q+4\eta_{5}\mathcal{Y}q)=0\nn\\
&&\partial_-\eta_{6}+\frac{1}{3}(4\eta_{2}p+2\eta_{7}p+3\eta_{5}\mathcal{L}p-12\eta_{8}q-8\eta_{1}\mathcal{L}q-4\eta_{6}\mathcal{L}q+24\eta_{5}\mathcal{W}q+\nn\\
&&\ \ \ \ \ \ \ \ \ \ \ 9\eta_{5}\mathcal{Y}p-8\eta_{1}\mathcal{Y}q-4\eta_{6}\mathcal{Y}q)=0\label{sl-}\\
&&\partial_-\eta_{7}+\frac{1}{3}(3\eta_{8}p-4\eta_{1}\mathcal{L}p+2\eta_{6}\mathcal{L}p-6\eta_{5}\mathcal{W}p+8\eta_2\mathcal{L}q-4\eta_{7}\mathcal{Y}q-12\eta_5\mathcal{L}^2q-\nn\\
&&\ \ \ \ \ \ \ \ \ \ \
8\eta_1\mathcal{W}q+16\eta_6\mathcal{W}q+6\eta_6\mathcal{Y}p+8\eta_2\mathcal{Y}q-4\eta_2\mathcal{Y}q-24\eta_5\mathcal{L}\mathcal{Y}q-12\eta_5\mathcal{Y}^2q)=0\nn\\
&&\partial_-\eta_8+\frac{1}{3}(-4\eta_2\mathcal{L}p+\eta_7\mathcal{L}p+4\eta_1\mathcal{W}p-2\eta_6\mathcal{W}p+4\eta_8\mathcal{L}q+8\eta_1\mathcal{L}^2q-4\eta_6\mathcal{L}^2q-8\eta_2\mathcal{W}q+\nn\\
&&\ \ \ \ \ \ \ \ \ \ \
8\eta_7\mathcal{W}q+3\eta_7\mathcal{Y}p+4\eta_8\mathcal{Y}q+16\eta_1\mathcal{L}\mathcal{Y}q-8\eta_6\mathcal{L}\mathcal{Y}q+8\eta_1\mathcal{Y}^2q-4\eta_6\mathcal{Y}^2q)=0\nn\\
&&\partial_-\eta_9+\frac{1}{3}(\eta_{10}p-4\eta_3p+4\eta_{11}q-8\eta_4q-4\eta_9\mathcal{L}q-4\eta_9\mathcal{Y}q)=0\nn\\
&&\partial_-\eta_{10}+\frac{1}{3}(2\eta_{11}p-4\eta_4p+3\eta_9\mathcal{L}p+12\eta_{12}q+4\eta_{10}\mathcal{L}q-8\eta_3\mathcal{L}q+24\eta_9\mathcal{W}q+9\eta_9\mathcal{Y}p+\nn\\
&&\ \ \ \ \ \ \ \ \ \ \ 4\eta_{10}\mathcal{Y}q-8\eta_3\mathcal{Y}q)=0\nn\\
&&\partial_-\eta_{11}+\frac{1}{3}(3\eta_{12}p+2\eta_{10}\mathcal{L}p+4\eta_3\mathcal{L}p+6\eta_9\mathcal{W}p+4\eta_{11}\mathcal{L}q+8\eta_4\mathcal{L}q+12\eta_9\mathcal{L}^2q+16\eta_{10}\mathcal{W}q+\nn\\
&&\ \ \ \ \ \ \ \ \ \ \
8\eta_3\mathcal{W}q+6\eta_{10}\mathcal{Y}p+4\eta_{11}\mathcal{Y}q+8\eta_4\mathcal{Y}q+24\eta_9\mathcal{L}\mathcal{Y}q+12\eta_9\mathcal{Y}^2q)=0\nn\\
&&\partial_-\eta_{12}+\frac{1}{3}(\eta_{11}\mathcal{L}p+4\eta_4\mathcal{L}p+2\eta_{10}\mathcal{W}p+4\eta_3\mathcal{W}p-4\eta_{12}\mathcal{L}q+4\eta_{10}\mathcal{L}^2q+8\eta_3\mathcal{L}^2q+8\eta_{11}\mathcal{W}q+\nn\\
&&\ \ \ \ \ \ \ \ \ \ \
8\eta_4\mathcal{W}q+3\eta_{11}\mathcal{Y}p-4\eta_{12}\mathcal{Y}q+8\eta_{10}\mathcal{L}\mathcal{Y}q+16\eta_3\mathcal{L}\mathcal{Y}q+4\eta_{10}\mathcal{Y}^2q+8\eta_3\mathcal{Y}^2q)=0\nn
\eea

Let us first consider the $BTZ$ black hole. This black hole is parameterized by\footnote{Here we have choose Branch 1 which could be the stable phase at the low temperature.}
\be
\mathcal{Y}=0,\ \mathcal{W}=0,\ q=0,\ p=0.
\ee
Then the situation is much like the case of $osp(3|2)$. The $``-''$ component equations lead to the conclusion that all $\eta_i$'s are independent of the $x^-$ coordinate. The $``+''$ component equations lead to
\bea
(\partial_+^2-\mathcal{L})\eta_1=0,\nn\\
(\partial_+^2-\mathcal{L})\eta_3=0,\nn\\
(\partial_+^2-\mathcal{L})(\partial_+^2-9\mathcal{L})\eta_5=0,\nn\\
(\partial_+^2-\mathcal{L})(\partial_+^2-9\mathcal{L})\eta_{9}=0.
\eea
As the analysis is similar to the $osp$ case, we just list the results as follows
%\begin{enumerate}
%\item
%For global AdS$_3$, $\mathcal{L}=\bar{\mathcal{L}}=-\frac{1}{4}$, there are (12,12) supersymmetry.
%\item
%For $\mathcal{L}=-\frac{1}{4}, \bar{\mathcal{L}}=-\frac{1}{9}$, there are
%\end{enumerate}
\begin{center}
\begin{tabular}{|c|c|c|}
\hline
Configuration&SUSY&Boundary Condition\\\hline
$\mathcal{L}=\bar{\mathcal{L}}=-\frac{1}{4}$&(12,12)&(AP,AP)\\\hline
$\mathcal{L}=-\frac{1}{4},\bar{\mathcal{L}}=-\frac{1}{9}$&(12,4)&(AP,P)\\\hline
$\mathcal{L}=-\frac{1}{4},\bar{\mathcal{L}}=-\frac{1}{36}$&(12,4)&(AP,AP)\\\hline
$\mathcal{L}=-\frac{1}{9},\bar{\mathcal{L}}=-\frac{1}{4}$&(4,12)&(P,AP)\\\hline
$\mathcal{L}=-\frac{1}{9},\bar{\mathcal{L}}=-\frac{1}{9}$&(4,4)&(P,P)\\\hline
$\mathcal{L}=-\frac{1}{9},\bar{\mathcal{L}}=-\frac{1}{36}$&(4,4)&(P,AP)\\\hline
$\mathcal{L}=-\frac{1}{36},\bar{\mathcal{L}}=-\frac{1}{4}$&(4,12)&(AP,AP)\\\hline
$\mathcal{L}=-\frac{1}{36},\bar{\mathcal{L}}=-\frac{1}{9}$&(4,4)&(AP,P)\\\hline
$\mathcal{L}=-\frac{1}{36},\bar{\mathcal{L}}=-\frac{1}{36}$&(4,4)&(AP,AP)\\\hline
\end{tabular}
\end{center}
Here in the table, the number in the bracket are the fermionic supercharges in the holomorphic and anti-holomorphic sectors respectively. The $AP$ or $P$
means the Killing spinor satisfy anti-periodic or periodic boundary condition in the $\phi$ direction. Besides the global AdS$_3$, the other solutions listed in the table are actually not smooth since their spatial holonomy is not in the center. Hence, they may be not allowed from the criteria of smoothness though they preserve some fermionic symmetry.

Moreover, one can also consider the massless BTZ and extreme BTZ black hole
\begin{enumerate}
\item For massless BTZ, $\mathcal{L}=\bar{\mathcal{L}}=0$, we have (4,4) supersymmetry.
\item For extreme BTZ with nonvanishing mass, $\mathcal{L}=0, \bar{\mathcal{L}}\not=0$ or $\mathcal{L}\not=0,\bar{\mathcal{L}}=0$, it preserves (4,0) or (0,4) supersymmetry.
\end{enumerate}

For $\mathcal{L}>0, \bar{\mathcal{L}}>0$, there could be supersymmetric configurations with non-vanishing higher spin charge. As the Killing spinor equations become complicated, we can restrict ourselves to constant spinor solutions.
The character matrix $M_+, M_-$ can be read from the equations (\ref{sl+}) and (\ref{sl-}). They are respectively
\begin{align}
M_+=\left(\begin{array}{cccccccccccc}
0&1&0&0&2\mathcal{Y}&0&0&0&0&0&0&0\\
\frac{1}{3}a&0&0&0&-2\mathcal{W}&\frac{2\mathcal{Y}}{3}&0&0&0&0&0&0\\
0&0&0&1&0&0&0&0&-2\mathcal{Y}&0&0&0\\
0&0&\frac{1}{3}a&0&0&0&0&0&-2\mathcal{W}&-\frac{2\mathcal{Y}}{3}&0&0\\
0&0&0&0&0&1&0&0&0&0&0&0\\
0&0&0&0&b&0&2&0&0&0&0&0\\
-\frac{4\mathcal{Y}}{3}&0&0&0&-2\mathcal{W}&\frac{2}{3}b&0&3&0&0&0&0\\
\frac{4\mathcal{W}}{3}&-\frac{4\mathcal{Y}}{3}&0&0&0&-\frac{2\mathcal{W}}{3}&\frac{1}{3}b&0&0&0&0&0\\
0&0&0&0&0&0&0&0&0&1&0&0\\
0&0&0&0&0&0&0&0&b&0&2&0\\
0&0&\frac{4\mathcal{Y}}{3}&0&0&0&0&0&2\mathcal{W}&\frac{2}{3}b&0&3\\
0&0&\frac{4\mathcal{W}}{3}&\frac{4\mathcal{Y}}{3}&0&0&0&0&0&\frac{2\mathcal{W}}{3}&\frac{1}{3}b&0
\end{array}
\right),\nn
\end{align}
\begin{align}
M_-=\left(\begin{array}{cccccccccccc}
0&\frac{5p}{3}&0&0&c_4&-\frac{1}{3}c_3&-\frac{2p}{3}&4q&0&0&0&0\\
\frac{1}{3}c_1&0&0&0&-c_5&\frac{1}{3}c_4&\frac{1}{3}c_3&-2p&0&0&0&0\\
0&0&0&\frac{5p}{3}&0&0&0&0&-c_4&-\frac{1}{3}c_3&\frac{2p}{3}&4q\\
0&0&\frac{1}{3}c_1&0&0&0&0&0&-c_5&-\frac{1}{3}c_4&\frac{1}{3}c_3&2p\\
\frac{4p}{3}&-\frac{8q}{3}&0&0&\frac{1}{3}c_3&\frac{p}{3}&-\frac{4q}{3}&0&0&0&0&0\\
-\frac{2}{3}c_3&\frac{4p}{3}&0&0&c_2&-\frac{1}{3}c_3&\frac{2p}{3}&-4q&0&0&0&0\\
-\frac{2}{3}c_4&\frac{2}{3}c_3&0&0&-c_5&\frac{2}{3}c_2&-\frac{1}{3}c_3&p&0&0&0&0\\
\frac{2}{3}c_5&-\frac{2}{3}c_4&0&0&0&-\frac{1}{3}c_5&\frac{1}{3}c_2&\frac{1}{3}c_3&0&0&0&0\\
0&0&-\frac{4p}{3}&-\frac{8q}{3}&0&0&0&0&-\frac{1}{3}c_3&\frac{p}{3}&\frac{4q}{3}&0\\
0&0&-\frac{2c_3}{3}&-\frac{4p}{3}&0&0&0&0&c_2&\frac{1}{3}c_3&\frac{2p}{3}&4q\\
0&0&\frac{2}{3}c_4&\frac{2}{3}c_3&0&0&0&0&c_5&\frac{2}{3}c_2&\frac{1}{3}c_3&p\\
0&0&\frac{2}{3}c_5&\frac{2}{3}c_4&0&0&0&0&0&\frac{1}{3}c_5&\frac{1}{3}c_2&-\frac{1}{3}c_3\\\end{array}
\right),\nn
\end{align}
in which some constants are defined as
\bea
&&a=3\mathcal{L}+5\mathcal{Y},\  b=3\mathcal{L}+\mathcal{Y},\ c_1=5\mathcal{L}p+16\mathcal{W}q+3\mathcal{Y}p,\  c_2=\mathcal{L}p+8\mathcal{W}q+3\mathcal{Y}p,\nn\\
&&
c_3=4(\mathcal{L}+\mathcal{Y})q,\ c_4=2(\mathcal{L}p+2\mathcal{W}q),\ c_5=2(\mathcal{W}p+2q(\mathcal{L}+\mathcal{Y})^2).
\eea
The determinant of $M_+$ and $M_-$ are respectively
\be
det(M_+)=(-9\mathcal{L}^3+16\mathcal{W}^2-21\mathcal{L}^2\mathcal{Y}+5\mathcal{L}\mathcal{Y}^2+25\mathcal{Y}^3)^2,\label{cha1}
\ee
\bea
det(M_-)&=&(\frac{1}{729}(729(-9\mathcal{L}^3+16\mathcal{W}^2-21\mathcal{L}^2\mathcal{Y}+5\mathcal{L}\mathcal{Y}^2+25\mathcal{Y}^3))p^6\nn\\
&&+3888\mathcal{W}q(5\mathcal{L}^2+46\mathcal{L}\mathcal{Y}+77\mathcal{Y}^2)p^5\nn\\
&&+5184q^2(46\mathcal{L}^4-9\mathcal{L}\mathcal{W}^3+208\mathcal{L}^3\mathcal{Y}+153\mathcal{W}^2\mathcal{Y}+372\mathcal{L}^2\mathcal{Y}^2+304\mathcal{L}\mathcal{Y}^3+94\mathcal{Y}^4)p^4\nn\\
&&+27648\mathcal{W}q^3(37\mathcal{L}^3+27\mathcal{W}^2+165\mathcal{L}^2\mathcal{Y}+219\mathcal{L}\mathcal{Y}^2+91\mathcal{Y}^3)p^3\nn\\
&&-36864q^4(\mathcal{L}+\mathcal{Y})^2(41\mathcal{L}^3-189\mathcal{W}^2+105\mathcal{L}^2\mathcal{Y}+87\mathcal{L}\mathcal{Y}^2+23\mathcal{Y}^3)p^2\nn\\
&&-442368\mathcal{W}q^5(\mathcal{L}+\mathcal{Y})(8\mathcal{L}^3-27\mathcal{W}^2+24\mathcal{L}^2\mathcal{Y}+24\mathcal{L}\mathcal{Y}^2+8\mathcal{Y}^3)p\nn\\
&&+16384q^6(8\mathcal{L}^3-27\mathcal{W}^2+24\mathcal{L}^2\mathcal{Y}+24\mathcal{L}\mathcal{Y}^2+8\mathcal{Y}^3)^2)^2. \label{cha2}
\eea
%Another interesting case is to have vanighing spin 3 charge and spin 3 chemical potential.
%\be
%\mathcal{W}=0,\ q=0
%\ee
The equations (\ref{sl+}) has non-zero constant solution if and only if
\be
-9\mathcal{L}^3+16\mathcal{W}^2-21\mathcal{L}^2\mathcal{Y}+5\mathcal{L}\mathcal{Y}^2+25\mathcal{Y}^3=0\label{con}
\ee
For general configuration, $\mathcal{L}\not=0,\mathcal{Y}\not=0,\mathcal{W}\not=0$, the equations (\ref{sl+}) can be solved by
\bea
&&\eta_1=C_1,\ \eta_2=-\frac{C_1 \mathcal{Y} (3 \mathcal{L} + 5 \mathcal{Y})}{3 \mathcal{W}},\ \eta_3=C_2,\nn\\
&&\eta_4=\frac{\eta_3 \mathcal{Y} (3 \mathcal{L} + 5 \mathcal{Y})}{3 \mathcal{W}},\ \eta_5=\frac{(3\mathcal{L}+5\mathcal{Y})C_1}{6\mathcal{W}},\ \eta_6=0\nn\\
&&\eta_7=-\frac{(3\mathcal{L}+\mathcal{Y})(3\mathcal{L}+5\mathcal{Y})C_1}{12\mathcal{W}},\ \eta_8=\frac{1}{3}C_1(\mathcal{L}+3\mathcal{Y}),\ \eta_9=\frac{(3\mathcal{L}+5\mathcal{Y})C_2}{6\mathcal{W}}\nn\\
&&\eta_{10}=0,\ \eta_{11}=-\frac{(3\mathcal{L}+\mathcal{Y})(3\mathcal{L}+5\mathcal{Y})C_2}{12\mathcal{W}},\ \eta_{12}=-\frac{1}{3}C_2(\mathcal{L}+3\mathcal{Y}).\label{kil}
\eea
 However the Killing spinor equations (\ref{sl-}) cannot be satisfied for the configuration (\ref{con}) and the solution (\ref{kil}). But for the extreme case $\bar{\tau}=\infty$, $q=p=0$, Eq. (\ref{sl-}) can be satisfied automatically. The condition $\bar{\tau}=\infty$ can be achieved by
\be
\bar{\mathcal{L}}=\bar{\mathcal{Y}},\ \mbox{or} \ \bar{\mathcal{W}}^2=\frac{16}{27}(\bar{\mathcal{L}}+\bar{\mathcal{Y}})^3.\label{con2}
\ee
 Therefore a general configuration (\ref{con}) and (\ref{con2}) preserve (2,0) supersymmetries. Note that we do not include the case $\bar{\mathcal{L}}=-\bar{\mathcal{Y}}$ as it leads to vanishing spin 3 charge $\bar{\mathcal{W}}=0$.

 There are some cases in which supersymmetry is enhanced. We consider the following cases.
\begin{enumerate}
\item $\mathcal{W}=0,\ q=0$. The spin 3 charge and chemical potential are zero, so that (\ref{cha1}) and (\ref{cha2}) reduce to
    \bea
   det(M_+)&=&(-9\mathcal{L}^3-21\mathcal{L}^2\mathcal{Y}+5\mathcal{L}\mathcal{Y}^2+25\mathcal{Y}^3)^2\nn\\
    &=&(\mathcal{L}-\mathcal{Y})^2(3\mathcal{L}+5\mathcal{Y})^4,\\
   det(M_-)&=&p^{12}(\mathcal{L}-\mathcal{Y})^2(3\mathcal{L}+5\mathcal{Y})^4.
    \eea
    The possible supersymmetric configurations are the following.
    \begin{enumerate}
    \item When $\mathcal{L}=-\frac{5}{3}\mathcal{Y}$,
         the solution of (\ref{sl+}) is
        \bea
        &&\eta_1=C_1,\ \eta_2=-2 C_3 \mathcal{Y},\ \eta_3=C_2,\ \eta_4=2 C_4\mathcal{Y},\ \eta_5=C_3,\ \eta_6=0\nn\\
        &&\eta_7=2 C_3 \mathcal{Y},\ \eta_8=\frac{4}{9}C_1\mathcal{Y},\ \eta_9=C_4,\ \eta_{10}=0,\ \eta_{11}=2 C_4\mathcal{Y},\ \eta_{12}=-\frac{4}{9}C_2\mathcal{Y}.\nn
        \eea
        There are four independent solutions, indicating four conserved supercharges. The equations (\ref{sl-}) can be satisfied by the extreme condition (\ref{con2}). Thus with the condition (\ref{con2}), the configuration preserve (4,0) supersymmetries in general.
    \item When $
    \mathcal{L}=\mathcal{Y}$,
    the solution of (\ref{sl+}) is
    \bea
    &&\eta_1=0,\ \eta_2=-2 C_1 \mathcal{Y},\ \eta_3=0,\ \eta_4=2C_2 \mathcal{Y},\ \eta_5=C_1,\ \eta_6=0\nn\\
    &&\eta_7=-2 C_1 \mathcal{Y},\ \eta_8=0,\ \eta_9=C_2,\ \eta_{10}=0,\ \eta_{11}=-2C_2 \mathcal{Y},\ \eta_{12}=0.\nn
    \eea
    There are two independent solutions, indicating two conserved supercharges. The equations (\ref{sl-}) can be satisfied in this case. Thus the configuration in this case in general  preserve (2,0) supersymmetries. Similar analysis in the anti-holomorphic sector suggests that the configuration $\mathcal{L}=\mathcal{Y},\bar{\mathcal{L}}=\bar{\mathcal{Y}}$ preserve (2,2) supersymmetries.
    \end{enumerate}
    In fact, the results here is similar to the ones in the $osp(3|2)$ case. This is expected as for $\mathcal{W}=\bar{\mathcal{W}}=0$, we can embed the configurations in $osp(3|2)$ gravity into the ones in $sl(3|2)$ gravity.

   \item  $\mathcal{Y}=0, p=0$. In this case, the relations (\ref{cha1}) and (\ref{cha2}) reduce to
       \bea
       det(M_+)&=&(9\mathcal{L}^3-16\mathcal{W}^2)^2,\nn\\
       det(M_-)&=&\frac{26835356}{531441}q^{12}(8\mathcal{L}^3-27\mathcal{W}^2)^4,\label{det}
       \eea
       which can only be satisfied by
       \be
       9\mathcal{L}^3-16\mathcal{W}^2=0,\ q=0
       \ee
   The condition $q=0$ can be achieved by extreme configuration (\ref{con2}). As the solution of equations (\ref{sl+}) is to set $\mathcal{Y}=0$ in the solution (\ref{kil}), this case is just a special limit of the general solution (\ref{kil}) and there is no more supersymmetry enhancement.
      \end{enumerate}
     % The most interesting configuration is the one that satisfies equations (\ref{sl+}) and (\ref{sl-}) without setting $q=p=0$.

\subsubsection{Supersymmetry II}

Now we analyze the condition from spatial holonomy to find the supersymmetric configurations.
%Note that the discussion of supersymmetry in $sl(3|2)$ is basically the same as we have done for $osp(3|2)$, besides the fact that there is a spin 3 hair.
For the solution we found in (\ref{sl32}),(\ref{sl32p}), we find the form of  $a_{\phi}$ to be
\begin{align}
a_{\phi}=\left(\begin{array}{cc}
a_{\phi}^{3\times3}&0\\0&a_{\phi}^{2\times2}\end{array}
\right).
\end{align}
It is a block diagonal matrix, with
\begin{align}
a_{\phi}^{3\times3}=\left(\begin{array}{ccc}
\frac{4}{3}q(\mathcal{L}+\mathcal{Y})&\sqrt{2} (\mathcal{L}+\mathcal{Y} -(\mathcal{L}+\mathcal{Y})p- 4 \mathcal{W} q )&-2(\mathcal{W}(-1+p)+2q(\mathcal{L}+\mathcal{Y}))\\
-\sqrt{2}(-1+p)&-\frac{8}{3}q(\mathcal{L}+\mathcal{Y})&\sqrt{2}(\mathcal{L}+\mathcal{Y} -(\mathcal{L}+\mathcal{Y})p- 4 \mathcal{W} q )\\
-4q&-\sqrt{2}(-1+p)&\frac{4}{3}q(\mathcal{L}+\mathcal{Y})
\end{array}
\right)\nn
\end{align}
and
\begin{align}
a_{\phi}^{2\times2}=\left(\begin{array}{cc}
0&(1+p)(\mathcal{L}-\mathcal{Y})\\
1+p&0\end{array}
\right)\nn\end{align}
We face a problem that the eigenvalues of the $3\times3$ matrix $a_{\phi}^{3\times3}$ are involved. However, the $2\times2$ matrix $a_{\phi}^{2\times2}$ is easy to deal with. Its eigenvalues turn out to be
\be
\varphi_{\bar{1}}=(1+p)\sqrt{\mathcal{L}-\mathcal{Y}},\ \varphi_{\bar{2}}=-(1+p)\sqrt{\mathcal{L}-\mathcal{Y}}.
\ee
Here we only consider the supersymmetric black holes for which
%which has a real entropy, which requires that $\mathcal{L}\geq\mathcal{Y}$. In other words, the eigenvalues of $a_{\phi}^{2\times2}$ are real.
%To have supersymmetry enhancement, it is required that one eigenvalue of $a_{\phi}^{2\times2}$ must be the same as one of
the eigenvalues of $a_{\phi}^{3\times3}$ have to be real. As the eigenvalues of $a_{\phi}^{2\times2}$ are real, then  $a_{\phi}^{3\times3}$ should satisfy the equation
\be
det(a_{\phi}^{3\times3}-x 1_{3\times3})=0
\ee
where $x=\varphi_{\bar{i}}$, with $\bar{i}=1$ or $2$. This leads to the equation
\bea
&&-128\mathcal{W}^2q^3-4\mathcal{W}(-1+p)(1+(-2+p)p+2q(-3x+8q(\mathcal{L}+\mathcal{Y})))\nn\\
&&+\frac{1}{27}(-3x+16q(\mathcal{L}+\mathcal{Y}))(64\mathcal{L}^2q^2-36(-1+p)^2\mathcal{Y}\nn\\
&&+(3x+8q\mathcal{Y})^2+4\mathcal{L}(-9-9(-2+p)p+4q(3x+8q\mathcal{Y})))=0\label{condition}
\eea
To check the consistency of (\ref{condition}), we can set $\mathcal{W}=q=0$ then we find the supersymmetry enhancement condition (\ref{susyenhance5})\footnote{There is some signature flipping due to the convention of the ansatz (\ref{sl32}).}. Therefore the analysis for the $\mathcal{W}$ vanishing case is the same as before.

The interesting case comes from $\mathcal{W}\not=0$. Though the condition is given in (\ref{condition}), the parameters $p,q$ are determined by (\ref{pq}),(\ref{tal}) and (\ref{entropyosp}). In general, there are solutions to (\ref{condition}), which could be quite involved. Here we would like to consider the constant solution and check the consistency with the discussion in the previous subsubsection.  We may choose $q=p=0$ by setting the antiholomorphic part to be extreme. Then the condition (\ref{condition}) simplifies to
\be
4\mathcal{W}+4(\mathcal{L}+\mathcal{Y})x-x^3=0.
\ee
It is just
\be
4\mathcal{W}=\pm\sqrt{\mathcal{L}-\mathcal{Y}}(3\mathcal{L}+5\mathcal{Y}),
\ee
which is the same as (\ref{con}).

\section{Discussion and Conclusion}

In this paper, we discussed the symmetries of the classical configurations in 3D higher spin gravity, defined by the Chern-Simons action. We found that the holonomy of the gauge potential around the spatial circle encodes the symmetry of the solution. To be modest, we only focused on two classes of configurations: the smooth one with maximal higher spin symmetries, and the higher spin black holes which may preserve part of supersymmetries.

We obtained the maximally symmetric solutions in various higher spin (super-)gravity theories and identified them with the smooth conical defects(surplus). For smooth conical defects(surplus), the spatial holonomies are in the centers of the corresponding gauge groups. Such configurations are of particular importance in the HS/CFT correspondence, as they correspond to the primary states $(0,\L_-)$ in the dual CFT. They are equally important as the global AdS$_3$, in the sense that they are also saddle points of the Euclidean path integral and could be taken as the vacua for different sectors. It is remarkable that the uniqueness of maximally symmetric space in the usual geometric sense is lost in 3D higher spin gravity. It would be interesting to investigate if the same phenomenon happens in the higher dimensional case.

The study of the maximally higher spin symmetric configurations may shed light on the
the HS/CFT correspondence in the semiclassical limit. In this limit, the central charge tends to infinity but the rank of gauge group $N$ is kept to be finite, the Vasiliev theory is simplified to a finite $sl(N)$ Chern-Simons gravity coupled to scalar matter. In \cite{Perlmutter:2012}, it was shown that the smooth conical surplus were in precise match with the primary states $(0,\L_-)$ in dual CFT, and the excitations on these surplus could be identified with more general primary states $(\L_+, \L_-)$. The match of the spectrum gives strong support of this duality, even though the theory becomes non-unitary. Our study on the smooth conical surplus  suggests that this duality could be true for the higher spin gravity theory for other gauge groups. We constructed the smooth conical defects for $Sp(2N), SO(2N+1), SO(2N), G_2$
and found precise match with the highest weight representations of $so(2N+1),sp(2 N),so(2N),g_2$. %This supports the  HS/CFT correspondence in the semi-classical limit for arbitrary semi-simple Lie algebra. However, this evidence is quite weak and much further study is deserved. It is necessary to make this correspondence more precise. One interesting investigation is to consider the fluctuations, both the higher spin ones and the scalar ones, around the smooth conical defects and compare these perturbative states with the states in the dual CFT.
This correspondence is captured by a simple relation:
\be
-i\th_i=\L_i+\rho_i,\nn
\ee
where $\th_i$'s are the eigenvalues of $a_+$, $\L_i$'s comprise the highest weight and $\rho_i$'s comprise the Weyl vector. It is interesting to see how this relation fits into some precise duality between AdS$_3$ higher spin gravity and coset minimal model.

In the higher spin supergravity theory, the picture is less clear. For the $sl(N+1|N)$ case, the smooth conical defects are indeed in match with the chiral primaries in the proposed dual CFT. However, for the $osp(2N+1|2N)$ case, the conical defects are not in agreement with the chiral primaries in the proposed supercoset. %The experience on the even spin minimal model holography tells us that there could be more than one holographic candidates for the even spin gravity. This could also be the case for the $osp(2N+1|2N)$ higher spin supergravity. 
 As the bosonic sector of $osp(2N+1|2N)$ involves both $so(2N+1)$ and $sp(2N)$ group, it is nontrivial for the dual CFT to match the spectrum of the smooth conical defects.

%Furthermore, our study on the $G_2$ case indicates that this duality could be true for arbitrary Lie algebra.

%There are two feature we need to emphasize. At first, it seems that the conical defects in $sp(2N)$ and $so(2N+1)$ are different due to the difference of the center of the group. So it would be interesting to see how to fit the even spin minimal model holography. Second, the conical defects(surplus)in $G_2$ gravity are in one-to-one correspondence with the representation of the $G_2$ group, indicating that the duality can be extended to $G_2$ gravity.

On the other hand, the smooth higher spin black holes are partially symmetric. They have trivial thermal holonomies in order to be smooth. But  they have nontrivial spatial holonomies, which allow us to analyze their symmetry properties. In general, they keep only the symmetries generated by the Cartan subalgebra of the gauge groups, suggesting that the constant solutions have well-defined global charges. For the black holes in the higher spin supergravity, the supersymmetric configurations are interesting.  For the higher spin supergravity, the Killing spinor should be generalized to account for the higher spin spinor. We focused on the supersymmetry of $osp(3|2),sl(3|2)$ higher spin black holes in this paper. We found that all the supersymmetric configurations with constant Killing spinor were extremal, but there were also non-extremal supersymmetric  configurations which have non-constant Killing spinors. This feature may hold for the  black holes in other higher spin super-gravity theories. We showed that it turned to be more efficient to work with the constraint (\ref{condf}) imposed by the spatial holonomy of the configuration to find the Killing spinors, even though it is possible to solve the Killing spinor equations directly.

For the black hole in the high spin supergravity discussed in the present work, their thermodynamics are relatively easy, due to the decoupling of the bosonic sectors. Their entropies are just the sum of the ones of the black holes in the decoupled theory. In this case, the entropies should be able to be understood from boundary CFT, as suggested in \cite{Gaberdiel:2012yb}. Even though the black holes in $osp(3|2)$ and $sl(3|2)$ look simple, their phase structure could be rich. In particular, the high temperature phase of $sl(3|2)$ black hole may present different features from usual spin 3 black hole, considering the fact that there could be two UV theories in this case\cite{Peng:2012ae}.

The supersymmetric black holes are of particular importance in string and supergravity, due to their better behavior under quantum corrections. In string theory, it has been found that for classes of BPS black holes, not only their entropy but also the quantum corrections are in exact agreement with the string prediction. Especially, the logarithmic corrections to the entropy function  due to massless modes provide  tests for the underlying quantum gravity\cite{Banerjee:2011jp,Sen:2011ba}. For the high spin supersymmetric black holes, it would be interesting to understand the possible logarithmic corrections to the entropy, coming from the massless graviton and high spin fields, from dual CFT.

%it is possible to define supersymmetry protected index, which is free of quantum correction. Therefore, the quantum corrections to the black hole quantities are much better under controlled. It would be interesting to study how to define this index in a high spin supergravity.

In our treatment of the black holes in $sl(3|2)$ gravity, we turned off the $U(1)$ field. It would be interesting to explore the black holes with non-vanishing $U(1)$ charge and search for the corresponding supersymmetric configurations. Besides, we showed for the first time that there were exact black hole solutions in higher spin super-gravity, whose entropy function could be written in an analytic form. It would be quite interesting to explore the thermodynamics of these higher spin black holes.

\vspace*{10mm}
\noindent {\large{\bf Acknowledgments}}\\
The work was in part supported by NSFC Grant No. 10975005, ~11275010.
\vspace*{5mm}

\section*{Appendix A:  $sl(N), sp(2N),so(2N+1),so(2N),g_2$}

Generally, the principal embedding of $sl(2)$ into a Lie algebra $g$ is defined to be the unique embedding with the number of $sl(2)$ modules equals to the rank of $g$. The spin of the modules in different Lie algebras is collected in the following table:
\begin{center}
\begin{tabular}{|c|c|}
\hline
$A_N$ & $2,3,\cdots,N+1$\\
\hline
$B_N$ & $2,4,\cdots,2N$\\
\hline
$C_N$ & $2,4,\cdots,2N$\\
\hline
$D_N$ & $2,4,\cdots,2N-2,N$\\
\hline
$E_6$ & $2,5,6,8,9,12$\\
\hline
$E_7$ & $2,6,8,10,12,14,18$\\
\hline
$E_8$ & $2,8,12,14,18,20,24,30$\\
\hline
$F_4$ & $2,6,8,12$\\
\hline
$G_2$ & $2,6$\\
\hline
\end{tabular}
\end{center}
These numbers equal to the degrees of different Casimir operators in the corresponding algebra.

For a Lie algebra with rank $r$, suppose we have a set of Chevalley basis $e_i^\pm,h_i$, which obeys the following commutation relations:
\be
[h_i,e_j^\pm]=\pm A_{ij}e_j^\pm,\ \ [e_i^+,e_j^-]=\delta_{ij}h_i
\ee
where $A_{ij}$ are elements of the Cartan matrix. Then in the principal embedding the generators $L_1,L_{-1}$ can be represented as:
\be
L_1=\sqrt{2}\sum_{i=1}^r e_i^+,L_{-1}=-\sqrt{2}\sum_{i=1}^r \sum_{j=1}^r A^{ij}e_i^-.
\ee
Here $A^{ij}=(A^{-1})_{ij}$. From the usual definition $L_0=[L_1,L_{-1}]/2$, one can easily check $[L_{\pm1},L_0]=\pm L_{\pm 1}$, hence $\{L_0,L_{\pm 1}\}$ indeed form a $sl(2)$ subalgebra.

Using this expression, as long as we have a matrix realization of the Chevalley basis and other lie algebra generators, we can construct the matrix realization of principal embedding in any Lie algebra.

In the case of $sl(N)$, we can use the set of $L_0,L_1,L_{-1}$ to generate the whole set of generators:
\begin{align}
&L_0=\frac{1}{2}\left(\begin{array}{ccccc}
N-1&0&\dots&0&0\\0&N-3&\dots&0&0\\ \vdots&\vdots&\ddots&\vdots&\vdots\\0&0&\dots&-(N-3)&0\\0&0&\dots&0&-(N-1)
\end{array}\right),\nonumber\\
&L_1=\left(
\begin{array}{ccccc}
0&0&\dots&0&0\\-\sqrt{N-1}&0&\dots&0&0\\0&-\sqrt{2N-4}&\dots&0&0\\ \vdots&\vdots&\ddots&\vdots&\vdots\\0&0&\dots&-\sqrt{N-1}&0
\end{array}
\right),\nonumber\\
&L_{-1}=\left(\begin{array}{ccccc}
0&\sqrt{N-1}&0&\dots&0\\0&0&\sqrt{2N-4}&\dots&0\\ \vdots&\vdots&\vdots&\ddots&\vdots\\0&0&0&\dots&\sqrt{N-1}\\0&0&0&\dots&0
\end{array}\right).
\end{align}
Here $L_{1(i+1,i)}=-\sqrt{Ni-i^2}$, $L_{-1(i,i+1)}=\sqrt{Ni-i^2}$, and they satisfy commutation relation $[L_i,L_j]=(i-j)L_{i+j}$.

The other generators are denoted by $W^s_m$, $3\leq s\leq N, -(s-1)\leq m\leq(s-1)$, and are given by
\begin{equation}
W^s_{s-1}=L_1^{s-1},\hs{2ex}W^s_{m-1}=-\frac{1}{s+m-1}[L_{-1},W^s_m].
\end{equation}
%We can explicitly multiply $N-1$ copies of $L_1$ to g

The generators of $sp(2N)$  can be obtained by truncating out odd spin generators from  $sl(2N)$. The generators of $so(2N+1)$ can be obtained by truncating out odd spin generators from $sl(2N+1)$. The generators of $g_2$ can be obtained by truncating out spin 4 generators from $so(7)$.

For the Lie algebra $so(2N)$, we use the set of generators $T_{ab}=-i(E_{ab}-E_{ba})$, which satisfying the following commutation relation:
\be
[T_{ab},T_{cd}]=-i(\delta_{bc} T_{ad}+\delta_{ad} T_{bc}-\delta_{bd} T_{ac}-\delta_{ac} T_{bd}).
\ee
The chavelley basis are
\bea
&& e_{l<N}^+=\frac{1}{2}(T_{2l,2l+1}-i T_{2l-1,2l+1}-i T_{2l,2l+2}-T_{2l-1,2l+2})\nn\\
&& e_{l<N}^-=\frac{1}{2}(T_{2l,2l+1}+i T_{2l-1,2l+1}+i T_{2l,2l+2}-T_{2l-1,2l+2})\nn\\
&& h_{l<N}=T_{2l-1,2l}-T_{2l+1,2l+2}\nn\\
&& e_N^+=\frac{1}{2}(T_{2N-2,2N-1}-i T_{2N-3,2N-1}+i T_{2N-2,2N}+T_{2N-3,2N})\nn\\
&& e_N^-=\frac{1}{2}(T_{2N-2,2N-1}+i T_{2N-3,2N-1}-i T_{2N-2,2N}+T_{2N-3,2N})\nn\\
&& h_N=T_{2N-3,2N-2}+T_{2N-1,2N}.
\eea

\section*{Appendix B: Superalgebra $sl(3|2)$ and $osp(3|2)$}

The generators of $sl(N|N-1)$ are classified into the bosonic ones
\be
W^{(1)s}_m(s=2,3,\cdots,N),\ W^{(2)s}_{m}(s=2,3,\cdots,N-1),\ J\nn\label{bosonic}
\ee
and the fermionic ones
\be
Q^{s}_{r}(s=1,\cdots,N),\ \bar{Q}^{s}_r(s=1,\cdots,N)\label{fermio}
\ee
where $-s+1\le m\le s-1$ and $-s+\frac{1}{2}\le r\le s-\frac{1}{2}$. One can find their realizations in \cite{Racah,Tan:2012}.
The generators of $osp(2N+1|2N)$ can be obtained by truncating out all the odd spin generators in (\ref{bosonic}) and one copy of the fermionic operators in  (\ref{fermio}). We  illustrate such kind of truncation below explicitly for the $sl(3|2)$ case, as we need supermatrix realization in the main context.

%\section*{Appendix B:Superalgebra }
Let us consider the $sl(3|2)$ higher spin supergravity. Its bosonic part is $sl(3)\oplus sl(2)\oplus u(1)$.
The generators are $L_i,A_i,W_m,J$, where $L_i,A_i$ are spin 2 generators and $W_m$ are spin 3 generators, $J$ is the $u(1)$ generator.
The fermionic part is generated by two spin $5/2$ and two spin $3/2$ generators, which are denoted as $S_r,T_r,G_s,H_s$ respectively. The commutation relations are
\bea
&&[L_i,L_j]=(i-j)L_{i+j},\ [A_i,A_j]=(i-j)L_{i+j},\ [L_i,A_j]=(i-j)A_{i+j},\nn\\
&&[L_i,W_m]=(2i-m)W_{i+m},\ [A_i,W_m]=(2i-m)W_{i+m},\nn\\
&&[W_m,W_n]=\frac{1}{6}(n-m)(2m^2+2n^2-m n-8)(L_{m+n}+A_{m+n}),\nn
\eea
\bea
&&[L_i,G_r]=(\frac{i}{2}-r)G_{i+r},\ [L_i,H_r]=(\frac{i}{2}-r)H_{i+r},\nn\\
&&[L_i,S_s]=(\frac{3i}{2}-s)S_{i+s},\ [L_i,T_s]=(\frac{3i}{2}-s)T_{i+s},\nn\\
&&[A_i,G_r]=\frac{4}{3}S_{i+r}+\frac{5}{3}(\frac{i}{2}-r)G_{i+r},\nn\\
&&[A_i,H_r]=-\frac{4}{3}T_{i+r}+\frac{5}{3}(\frac{i}{2}-r)H_{i+r},\nn\\
&&[A_i,S_s]=\frac{1}{3}(\frac{3i}{2}-s)S_{i+s}-\frac{1}{3}(3i^2-2is+s^2-\frac{9}{4})G_{i+s},\nn\\
&&[A_i,T_s]=\frac{1}{3}(\frac{3i}{2}-s)T_{i+s}+\frac{1}{3}(3i^2-2is+s^2-\frac{9}{4})H_{i+s},\nn
\eea
\bea
&&[W_m,G_r]=-\frac{4}{3}(\frac{m}{2}-2r)S_{m+r},\ [W_m,H_r]=-\frac{4}{3}(\frac{m}{2}-2r)T_{m+r},\nn\\
&&[W_m,S_s]=-\frac{1}{3}(2s^2-2sm+m^2-\frac{5}{2})S_{m+s}-\frac{1}{6}(4s^3-3s^2m+2sm^2-m^3-9s+\frac{19}{4}m)G_{m+s},\nn\\
&&[W_m,T_s]=\frac{1}{3}(2s^2-2sm+m^2-\frac{5}{2})T_{m+s}-\frac{1}{6}(4s^3-3s^2m+2sm^2-m^3-9s+\frac{19}{4}m)H_{m+s},\nn
\eea
\bea
&&[J,L_i]=0,\ [J,A_i]=0,\ [J,W_m]=0,\ [J,G_r]=G_r,\ [J,H_r]=-H_r,\nn\\
&&[J,S_r]=S_r,\ [J,T_r]=-T_r,\ \{G_r,G_s\}=0,\ \{H_r,H_s\}=0,\nn\\ &&\{S_r,S_s\}=0,\ \{T_r,T_s\}=0,\ \{G_r,S_s\}=0,\ \{H_r,T_s\}=0,\nn\\
&&\{G_r,H_s\}=2L_{r+s}+(r-s)J,\nn\\
&&\{S_r,T_s\}=-\frac{3}{4}(r-s)W_{r+s}+\frac{1}{8}(3s^2-4rs+3r^2-\frac{9}{2})(L_{r+s}-3A_{r+s})-\frac{1}{4}(r-s)(r^2+s^2-\frac{5}{2})J,\nn\\
&&\{G_r,T_s\}=-\frac{3}{2}W_{r+s}+\frac{3}{4}(3r-s)A_{r+s}-\frac{5}{4}(3r-s)L_{r+s},\nn\\
&&\{H_r,S_s\}=-\frac{3}{2}W_{r+s}-\frac{3}{4}(3r-s)A_{r+s}+\frac{5}{4}(3r-s)L_{r+s}\nn
\eea
Note that one can truncate out the spin 3 and spin 1 generators and one copy of spin $3/2$ and $5/2$ generators. The resulting algebra is just $osp(3|2)$. More explicitly, one can define the remaining spin $5/2$ and spin $3/2$ generators as $R_s$ and $Z_r$, which are
\be
R_s=S_s-T_s,\ Z_r=G_r+H_r.
\ee
One can show that the subalgebra $\{L_i,A_i,R_s,Z_r\}$ is closed. The commutation relations are
\bea
&&[L_i,L_j]=(i-j)L_{i+j},\ [A_i,A_j]=(i-j)L_{i+j},\ [L_i,A_j]=(i-j)A_{i+j},\nn\\
&&[L_i,R_r]=(\frac{3i}{2}-r)R_{i+r},\ [L_i,Z_r]=(\frac{i}{2}-r)Z_{i+r}\nn\\
&&[A_i,R_r]=\frac{1}{3}(\frac{3i}{2}-r)R_{i+r}-\frac{1}{3}(3i^2-2ir+r^2-\frac{9}{4})Z_{i+r}\nn\\
&&[A_i,Z_r]=\frac{4}{3}R_{i+r}+\frac{5}{3}(\frac{i}{2}-r)Z_{i+r}\nn\\
&&\{R_r,R_s\}=-\frac{1}{4}(3r^2-4rs+3s^2-\frac{9}{2})(L_{r+s}-3A_{r+s})\nn\\
&&\{Z_r,Z_s\}=4L_{r+s},\nn\\
&&\{R_r,Z_s\}=-\frac{3}{2}(3s-r)A_{r+s}+\frac{5}{2}(3s-r)L_{r+s}\nn
\eea
The matrix realization of the $sl(3|2)$ generators are
\begin{align}
&L_0=\left(\begin{array}{ccccc}
1&0&0&0&0\\0&0&0&0&0\\0&0&-1&0&0\\0&0&0&\frac{1}{2}&0\\0&0&0&0&-\frac{1}{2}
\end{array}\right),\ L_1=\left(
\begin{array}{ccccc}
0&0&0&0&0\\ \sqrt{2}&0&0&0&0\\0&\sqrt{2}&0&0&0\\0&0&0&0&0\\0&0&0&1&0
\end{array}
\right)\nonumber\\
&L_{-1}=\left(\begin{array}{ccccc}
0&-\sqrt{2}&0&0&0\\0&0&-\sqrt{2}&0&0\\ 0&0&0&0&0\\0&0&0&0&-1\\0&0&0&0&0
\end{array}\right),\ A_0=\left(\begin{array}{ccccc}
1&0&0&0&0\\0&0&0&0&\\0&0&-1&0&0\\0&0&0&-\frac{1}{2}&0\\0&0&0&0&\frac{1}{2}\end{array}
\right)\nonumber\\
&A_1=\left(\begin{array}{ccccc}
0&0&0&0&0\\ \sqrt{2}&0&0&0&0\\0&\sqrt{2}&0&0&0\\0&0&0&0&0\\0&0&0&-1&0\end{array}
\right)\ A_{-1}=\left(\begin{array}{ccccc}
0&-\sqrt{2}&0&0&0\\0&0&-\sqrt{2}&0&0\\0&0&0&0&0\\0&0&0&0&1\\0&0&0&0&0\end{array}
\right)\nonumber\\
&W_2=\left(\begin{array}{ccccc}
0&0&0&0&0\\ 0&0&0&0&0\\4&0&0&0&0\\0&0&0&0&0\\0&0&0&0&0\end{array}
\right)\ W_1=\left(\begin{array}{ccccc}
0&0&0&0&0\\ \sqrt{2}&0&0&0&0\\0&-\sqrt{2}&0&0&0\\0&0&0&0&0\\0&0&0&0&0\end{array}
\right)\nonumber
\end{align}
\begin{align}
&W_0=\left(\begin{array}{ccccc}
\frac{2}{3}&0&0&0&0\\ 0&-\frac{4}{3}&0&0&0\\0&0&\frac{2}{3}&0&0\\0&0&0&0&0\\0&0&0&0&0\end{array}
\right), W_{-1}=\left(\begin{array}{ccccc}
0&-\sqrt{2}&0&0&0\\ 0&0&\sqrt{2}&0&0\\0&0&0&0&0\\0&0&0&0&0\\0&0&0&0&0\end{array}
\right)\nonumber\\
&W_{-2}=\left(\begin{array}{ccccc}
0&0&4&0&0\\ 0&0&0&0&0\\0&0&0&0&0\\0&0&0&0&0\\0&0&0&0&0\end{array}
\right)\ J=\left(\begin{array}{ccccc}
2&0&0&0&0\\ 0&2&0&0&0\\0&0&2&0&0\\0&0&0&3&0\\0&0&0&0&3\end{array}
\right)\nonumber
\end{align}
\begin{align}
&G_{\frac{1}{2}}=\left(\begin{array}{ccccc}
0&0&0&0&0\\ 0&0&0&0&0\\0&0&0&0&0\\ 2&0&0&0&0\\0&\sqrt{2}&0&0&0\end{array}
\right)\ G_{-\frac{1}{2}}=\left(\begin{array}{ccccc}
0&0&0&0&0\\ 0&0&0&0&0\\0&0&0&0&0\\0&-\sqrt{2}&0&0&0\\0&0&-2&0&0\end{array}
\right)\nonumber\\
&H_{\frac{1}{2}}=\left(\begin{array}{ccccc}
0&0&0&0&0\\ 0&0&0&\sqrt{2}&0\\0&0&0&0&2\\0&0&0&0&0\\0&0&0&0&0\end{array}
\right)\ H_{-\frac{1}{2}}=\left(\begin{array}{ccccc}
0&0&0&2&0\\ 0&0&0&0&\sqrt{2}\\0&0&0&0&0\\0&0&0&0&0\\0&0&0&0&0\end{array}
\right)\nn
\end{align}
\begin{align}
&S_{\frac{3}{2}}=\left(\begin{array}{ccccc}
0&0&0&0&0\\ 0&0&0&0&0\\0&0&0&0&0\\0&0&0&0&0\\-3&0&0&0&0\end{array}
\right)\ S_{\frac{1}{2}}=\left(\begin{array}{ccccc}
0&0&0&0&0\\ 0&0&0&0&0\\0&0&0&0&0\\-1&0&0&0&0\\0&\sqrt{2}&0&0&0\end{array}
\right)\nonumber\\
&S_{-\frac{1}{2}}=\left(\begin{array}{ccccc}
0&0&0&0&0\\ 0&0&0&0&0\\0&0&0&0&0\\0&\sqrt{2}&0&0&0\\0&0&-1&0&0\end{array}
\right)\ S_{-\frac{3}{2}}=\left(\begin{array}{ccccc}
0&0&0&0&0\\ 0&0&0&0&0\\0&0&0&0&0\\0&0&-3&0&0\\0&0&0&0&0\end{array}
\right)\nonumber
\end{align}
\begin{align}
&T_{\frac{3}{2}}=\left(\begin{array}{ccccc}
0&0&0&0&0\\ 0&0&0&0&0\\0&0&0&-3&0\\0&0&0&0&0\\0&0&0&0&0\end{array}
\right)\ T_{\frac{1}{2}}=\left(\begin{array}{ccccc}
0&0&0&0&0\\ 0&0&0&-\sqrt{2}&0\\0&0&0&0&1\\0&0&0&0&0\\0&0&0&0&0\end{array}
\right)\nonumber\\
&T_{-\frac{1}{2}}=\left(\begin{array}{ccccc}
0&0&0&-1&0\\ 0&0&0&0&\sqrt{2}\\0&0&0&0&0\\0&0&0&0&0\\0&0&0&0&0\end{array}
\right)\ T_{-\frac{3}{2}}=\left(\begin{array}{ccccc}
0&0&0&0&3\\ 0&0&0&0&0\\0&0&0&0&0\\0&0&0&0&0\\0&0&0&0&0\end{array}
\right)\nn
\end{align}
The fermionic generators of $osp(3|2)$ can be realized by
\begin{align}
&R_{\frac{3}{2}}=\left(\begin{array}{ccccc}
0&0&0&0&0\\ 0&0&0&0&0\\0&0&0&3&0\\0&0&0&0&0\\-3&0&0&0&0\end{array}
\right)\ R_{\frac{1}{2}}=\left(\begin{array}{ccccc}
0&0&0&0&0\\ 0&0&0&\sqrt{2}&0\\0&0&0&0&-1\\-1&0&0&0&0\\0&\sqrt{2}&0&0&0\end{array}
\right)\nonumber\\
&R_{-\frac{1}{2}}=\left(\begin{array}{ccccc}
0&0&0&1&0\\ 0&0&0&0&-\sqrt{2}\\0&0&0&0&0\\0&\sqrt{2}&0&0&0\\0&0&-1&0&0\end{array}
\right)\ R_{-\frac{3}{2}}=\left(\begin{array}{ccccc}
0&0&0&0&-3\\ 0&0&0&0&0\\0&0&0&0&0\\0&0&-3&0&0\\0&0&0&0&0\end{array}
\right)\nn\\
&Z_{\frac{1}{2}}=\left(\begin{array}{ccccc}
0&0&0&0&0\\ 0&0&0&\sqrt{2}&0\\0&0&0&0&2\\2&0&0&0&0\\0&\sqrt{2}&0&0&0\end{array}
\right)\ Z_{-\frac{1}{2}}=\left(\begin{array}{ccccc}
0&0&0&2&0\\ 0&0&0&0&\sqrt{2}\\0&0&0&0&0\\0&-\sqrt{2}&0&0&0\\0&0&-2&0&0\end{array}
\right)\nn
\end{align}
After some redefinition of the matrix we find that the commutation relations are the same as \cite{Justin:susy}. Our choice is to match the bosonic commutation relations given in \cite{Tan:2012}.

\end{document}